\documentclass[12pt, draftclsnofoot,onecolumn]{IEEEtran}
\usepackage{tabularx}
\usepackage{ifpdf}
\usepackage{graphicx}
\usepackage{amssymb}
\usepackage{amsmath}
\usepackage{cite}
\usepackage{epstopdf}
\usepackage{mathrsfs}
\usepackage{cite}
\usepackage[dvipsnames]{xcolor}
\usepackage[keeplastbox]{flushend}

\newtheorem{corollary}{Corollary}

\newcommand{\xd}{\nonumber \\}
\newcommand{\PDF}[2]{f_{{#1}}\left({#2}\right)}
\newcommand{\CDF}[2]{F_{{#1}}\left({#2}\right)}

\newcommand{\Ei}{\mathrm{Ei}}

\begin{document}
\title{Secrecy Performance of Small-Cell Networks with Transmitter Selection and Unreliable Backhaul under Spectrum Sharing Environment}
\author{Jinghua Zhang, Chinmoy Kundu,  Octavia A. Dobre, Emi Garcia-Palacios, and Nguyen-Son Vo
\thanks{J. Zhang, and E. Garcia-Palacios are with Queen's University Belfast, UK (email: jzhang22@qub.ac.uk and e.garcia@ee.qub.ac.uk).}
}
\author{\IEEEauthorblockN{Jinghua Zhang, Chinmoy Kundu, Octavia A. Dobre, Emiliano Garcia-Palacios, and Nguyen-Son Vo }



        \thanks{Part of this work was presented in the 13th EAI International Conference on Cognitive Radio Oriented Wireless Networks (CROWNCOM), Ghent, Belgium, Sep. 2018.}

        \thanks{Jinghua Zhang and Emiliano Garcia-Palacios are with the School of Electronics, Electrical Engineering and Computer Science, Queen's University Belfast, Northern Ireland, U.K. (e-mail: jzhang22@qub.ac.uk and e.garcia@ee.qub.ac.uk.)}

 \thanks{Chinmoy Kundu is with Indian Institute of Technology Jammu. This work was completed in parts when he was with 
 Queen's University Belfast and subsequently with 
 University of Texas at Dallas, USA. (e-mail: chinmoy.kundu@iitjammu.ac.in)}
 
  \thanks{Octavia A. Dobre is with the Engineering and Applied Science, Memorial University, Canada. (e-mail: odobre@mun.ca.)}
 
 \thanks{Nguyen-Son Vo is with the Institute of Fundamental and Applied Sciences,
Duy Tan University, Vietnam. (e-mail: vonguyenson@dtu.edu.vn.)}

\thanks{This work was supported in part by the Royal Society-SERB Newton International Fellowship under Grant NF151345 and Natural Science and Engineering Research Council of Canada (NSERC) through
its Discovery program.} 
}
\maketitle

\begin{abstract}
We investigate the secrecy performance of an underlay small-cell cognitive radio network under unreliable 
backhaul connections. The small-cell network shares the same spectrum with the primary network, ensuring that a desired outage 
probability constraint is always met in the primary network. 
{To improve the security of the small-cell cognitive network, we  propose three sub-optimal small-cell transmitter selection schemes,} 
namely sub-optimal transmitter selection, minimal interference selection, and minimal eavesdropping selection.  Closed-form expressions of the non-zero secrecy rate, 
secrecy outage probability, and ergodic secrecy capacity are provided for the  schemes along with asymptotic expressions.
{We also propose an optimal selection scheme and compare performances with the sub-optimal selection schemes.} 
{Computable expressions for the non-zero secrecy rate and  
secrecy outage probability are presented for the optimal selection scheme.}
Our results show that by increasing the primary transmitter's power and the number of small-cell transmitters, the system performance improves. 
The selection scheme, the backhaul reliability, and the primary user quality-of-service constraint also have a significant impact on secrecy performance.

\end{abstract}
\section{Introduction}
The Internet-of-Things (IoT) paradigm has been driving an explosion in the deployment of wireless networks, which will be highly dense and heterogeneous \cite{Jeffrey20145G}. To enable dense wireless networks and achieve high data rates,  backhaul links between macro-cells and small-cells are needed to provide broadband communication in heterogeneous networks (HetNets) \cite{chia2009next}. The traditional wired backhaul network has shown high reliability and high data rate, however, the cost of deploying and sustaining the large-scale wired links require huge investment {\cite{Orawan2011Evolution,Xiaohu20145G, coldrey2012small}}. Hence, wireless backhaul has emerged as a suitable alternative solution, as being cost-effective and flexible in practical systems.  {Wireless backhaul is intrinsically unreliable  due to non-line-of-sight (n-LOS) propagation and vulnerability to fading. Enhancing the reliability of wireless backhaul has resulted in further investigation\cite{kim2017performance,khan2015performance}.}

The need to deploy new wireless systems is also pushing the demand for frequency resources. 
A report released by the Federal Communications Commission (FCC) showed that most of licensed spectrum bands are occupied 
\cite{Kolodzy2002spectrum}. Cognitive radio (CR) is an effective solution to overcome the inefficient frequency usage and 
spectrum scarcity  \cite{Mitola1999Cognitive}. CR allows unlicensed secondary users to dynamically and opportunistically access 
the licensed spectrum of the primary users.  {  A CR underlay scheme is a possible way to ensure unlicensed secondary 
users and licensed primary users can transmit information concurrently. Concurrent transmission can happen as long as the interference at the 
licensed primary users does not exceed a threshold.} The primary network is guaranteed reliable communication in 
the presence of the secondary network.

In \cite{zhong2011outage}, the authors studied the outage probability of an 
underlay cognitive network deploying decode-and-forward relaying in Nakagami-\textit{m} fading channels with the 
interference temperature constraint. 
Optimization of time and power allocation in the secondary network is studied in \cite{Lee2015Cognitive}. 
A survey on the radio resource allocation techniques for efficient spectrum access in CR networks can be found 
in \cite{octavia2016survey}.
 {More recently, 
the authors in \cite{sultana2017efficient} optimized the resource allocation of a device-to-device system 
using the cognitive approach. In this case, device-to-device users opportunistically
access the underutilized radio spectrum.} In \cite{Zhangjinghua}, the impact of the primary network on the secondary network is investigated when the primary transmitter is close to the secondary transmitter or further away. 
Impacts of unreliable backhaul links on the cooperative HetNets in spectrum sharing environment is studied in \cite{huy2017cognitive}. 

Network performance can be enhanced by improving diversity through selection, which is considered in this paper. Examples of relay selection in non-cognitive and secondary user selection in cognitive radio networks are found in 
\cite{Trung2013Cognitive,Kundu2016Relay,jinghuazhang2017Cognitive,nguyen2017multiuser,nguyen2017cognitive}. 
The best relay selection is explored in cognitive amplify-and-forward relay system in \cite{Trung2013Cognitive}.
The authors in \cite{Kundu2016Relay} investigated various relay selection schemes to enhance the security of a cooperative system with threshold-selection decode-and-forward relays. 
In \cite{jinghuazhang2017Cognitive}, the authors considered secondary user selection schemes to enhance the outage probability of the secondary network. Later, various relay selection schemes and multiuser scheduling are studied while considering unreliable backhaul links in cognitive networks in \cite{nguyen2017multiuser}. Performance of the best relay selection scheme is explored in cognitive HetNets along with unreliable backhaul connections in \cite{nguyen2017cognitive}.  

Due to the broadcast nature of the wireless channels, information in a wireless network is vulnerable to wiretapping.  { In practice, CR networks are more susceptible to eavesdropping due to the dynamic and open nature of the network architecture. 
Researchers have studied physical layer security (PLS) in cooperative and cognitive radio network extensively.
To begin with, the secrecy performance of an energy harvesting relay system was analyzed in \cite{Phong2016Secure}.} Multiple trusted relays in the communication between a legitimate source and a legitimate destination are considered.
In \cite{nguyen2016joint}, the authors studied the secrecy rate maximization problem for multiple-input single-output broadcast channel in the presence of multiple
passive eavesdroppers and primary users.
Authors in \cite{kundu2016secrecy} investigated secrecy outage probability and ergodic secrecy rate of a threshold-selection decode-and-forward relay system. 
In \cite{Yuzhen2016Secure} and \cite{Kundu2017AFrelay}, the secrecy outage performance of dual-hop amplify-and-forward  relaying systems was investigated. 


 {In \cite{liu2016divide, liu2016cooperative, liu2016jammer}}, {the authors introduced a divide-and-conquer based method in which the source message is encoded in multiple blocks and transmitted separately one-by-one. }
  {This  method can secure communication from multiple eavesdroppers without knowing the eavesdropper's location in the region. Secrecy performance is analyzed with the help of stochastic geometry. Following this work, in \cite{liu2016jammer}, the authors investigated the jammer placement method to approximately utilize the minimum number of jammers.
Contrary to the general studies in stochastic geometry, the authors in \cite{liu2017physical} consider a practical scenario of finite area consisting of a transmitter, a legitimate receiver and several eavesdroppers.
Covert communication can prevent adversaries from knowing that a communication has actually occurred. 
The authors in \cite{liu2018covert} investigated convert wireless communication utilizing interference available in the network. Performance is analyzed using stochastic geometry.}

Considering unreliable wireless backhaul connections, the authors in \cite{nguyen2017secure} investigated the secrecy performance of cooperative single carrier HetNets.  {The authors in \cite{Yincheng} studied the effect of unreliable wireless backhaul on the secrecy performance of an energy harvesting relay network. Different transmitter selection schemes were proposed to enhance the secrecy.} The secrecy performance of finite-sized cooperative single carrier systems with multiple passive eavesdroppers and unreliable wireless backhaul connections is explored in \cite{kim2016secrecy}. Wireless backhaul along with secrecy is only recently being considered for CR network in \cite{Vu2017Secure}.

Nevertheless, the aforementioned works did not take into account the impact of unreliable backhaul on PLS in an ``integrated," i.e. more realistic, CR network. Other important aspects of such network were missing. For example, \cite{nguyen2017multiuser,huy2017cognitive,Vu2017Secure,nguyen2017cognitive} only considered interference on the primary network and neglected  interference at the secondary network from primary users. Backhaul reliability was considered for PLS in \cite{Yincheng,kim2016secrecy,nguyen2017secure}, however, not in CR networks. Though \cite{Vu2017Secure} took into account the wireless backhaul reliability for security in CR network, the system model is different from ours.      {Interference from the primary network both at the secondary eavesdropper and destination were not considered. This makes performance evaluation easier.}
Neither non-zero secrecy rate nor ergodic secrecy rate were derived. User selection schemes were also different. The impact of guaranteeing the QoS in the primary network was not considered either in the aforementioned research. These are important considerations that we take into account in this paper.

 {Our research addresses these key issues in CR networks with wireless backhaul along with power consumption and complexity. We consider a CR network with a secondary macro-cell base station (BS) connected to multiple secondary small-cell transmitters via unreliable wireless backhaul. To improve the secrecy of the secondary network and reduce the power consumption and complexity of the system, we propose three sub-optimal small-cell transmitter selection schemes,} {namely: (i) sub-optimal transmitter selection (STS), which maximizes the channel gain from the secondary source to the destination; (ii) minimal interference selection (MIS), which minimizes the channel gain from the secondary source to the primary receiver; (iii) minimal eavesdropping selection (MES), which minimizes the channel gain from the primary source to the eavesdropper. An optimal selection (OS) scheme is also proposed, in which secondary secrecy capacity reaches maximum.} 

Our contributions in this paper are summarised as follows:
\begin{enumerate}
\item We take into account the backhaul reliability when calculating the secrecy performance, to show its important impact. 
\item Different from existing literature, we consider both  interference at the primary network from the secondary network and at the secondary network from the primary network.
\item We consider the outage probability as a primary QoS constraint metric, which is different from previous works in CR.

\item  {
We derive the closed-form and asymptotic expressions of the non-zero secrecy rate, secrecy outage probability (SOP), and ergodic secrecy capacity for the sub-optimal selections schemes. In addition, we present a computable formula for the non-zero secrecy rate and SOP for the OS scheme.}

\item Finally, we also assess the impact of varying the number of small-cell 
transmitters upon the system performance. 

\end{enumerate}

The rest of the paper is organised as follows. In Section \ref{section 2}, the system and channel models are described. The selection schemes at the small-cell transmitters are introduced in Section \ref{section 3}. The non-zero secrecy rate, secrecy outage probability and ergodic secrecy rate of the proposed system are demonstrated in Section \ref{section 4}, \ref{section 5}, and \ref{section 6}, respectively. The asymptotic secrecy analyses are formulated in Section \ref{section 7}. Numerical results from Monte-Carlo simulations are showcased in Section \ref{section 8}. Finally, the paper is concluded in Section \ref{section 9}.   

\textit{Notation:} $\mathbb{P}[\cdot]$ is the probability of occurrence of an event. 
For a random variable $X$,  $\mathbb{E}_X[\cdot]$ denotes expectation or mean of 
$X$, $F_{XY} (\cdot)$ denotes the cumulative distribution function (CDF) of the 
signal-to-interference-plus-noise-ratio (SINR) between nodes $X$-$Y$, $\Gamma_{XY}$,  and 
$f_{XY} (\cdot)$ represents the corresponding probability density function (PDF). 

\section{System and channel model}\label{section 2}
\begin{figure}
 \centering 
 \includegraphics[width=2.3in]{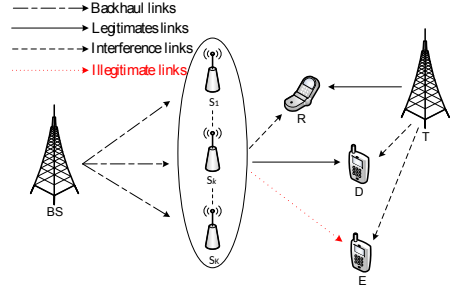} 
 \caption{Underlay cognitive radio network with unreliable backhaul connections.}
 \label{fig:SM}
 \end{figure}
{As depicted in Fig. \ref{fig:SM}, the system consists of a primary network and a cognitive secondary network, which shares the spectrum of the primary network. The primary network includes a primary transmitter, 
$T$, and its receiver, $R$. 
The secondary network is assumed to be a dense heterogeneous network in which a macro cell BS provides unreliable wireless backhauls to $K$ small-cell transmitters, $S_k$, $ \forall k=1 \cdots, K$. The small-cell transmitters in turn serving an end-user $D$. A cooperative one-shot communications like in \cite{kim2017performance} is considered, where if the message does not arrive from the BS through backhaul within a reasonable time, a transmitter does not transmit. No coding, modulation, automatic repeat
request, or power control are applied when a backhaul transmission fails.}
An eavesdropper $E$ is assumed to intercept 
the secondary transmission \footnote{This paper considers secrecy of a single secondary destination and eavesdropper case; however, an interesting future direction would be to analyze secrecy for multiple destinations and eavesdroppers.}. As both primary and secondary transmissions happen concurrently, interference 
from the primary affects the secondary reception at $D$ and $E$; likewise, secondary transmission affects the primary receiver $R$.   { This system model can be found in future dense heterogeneous networks. The BS may be a typical cellular transmitter providing backhaul connectivity to the small-cell BSs, which in turn provide wireless coverage to an indoor destination user \cite{khan2015performance, kim2017performance, kim2016secrecy}.}

The channel gains between nodes are assumed to be independent Rayleigh distributed 
and their power gains are exponentially distributed. Power gains of $S_k$-$D$,  $S_k$-$E$ and $S_k$-$R$ for all $k$ are independently distributed with exponential parameters $\lambda_{sd}$, $\lambda_{se}$, and $\lambda_{sr}$, 
respectively.  {It is to be noted here that $\lambda_{sd}$, $\lambda_{se}$, and $\lambda_{sr}$ may not be identical. Independent and identically distributed channel conditions separately for $S_k$-$D$, $S_k$-$E$, and $S_k$-$R$ may arise in cases when a cluster of small-cell transmitters serves a destination \cite{khan2015performance, kim2016secrecy}}.
The approach in this paper can however be extend to the independent not identically distributed cases by following \cite{Kundu_relsel} as well.

Noise at receivers is modeled as additive white Gaussian noise (AWGN) with zero mean and variance $N_0$.
To mitigate the effect of eavesdropping on secondary transmission, the best transmitter is 
selected among all transmitters (using proposed rules) to transfer the information to $D$.

While messaging from the macro BS to small-cell transmitters, the wireless backhaul link might have certain probability of failure. The backhaul reliability is modeled as a Bernoulli process, $\mathbb{I}_k$ for $k=1,\cdots, K$, with success probability $\mathbb{P}(\mathbb{I}_k=1)=\Lambda$, and failure probability 
$\mathbb{P}(\mathbb{I}_k=0)=1-\Lambda$,  $ \forall k=1,\cdots, K$. {This alternatively means that a small-cell transmitter may fail to deliver its message with probability (1-$\Lambda$) due to its backhaul failure.}

 {To improve the security of the secondary CR network, three sub-optimal selection rules are proposed along with an optimal selection rule. These selection schemes utilize the diversity benefit of multiple secondary transmitters depending on the available channel knowledge.} {The optimal selection scheme requires full knowledge of all channels, whereas only a single channel knowledge is required for sub-optimal selections.}   The non-zero secrecy rate, SOP, and ergodic secrecy rate are derived as a measure of secrecy of the secondary network  for the proposed transmitter selection schemes described in details later. Diversity analysis is also provided for each performance metric to obtain critical insight in high SNR regime. 



 {\subsection{Secondary Transmit Power Constraint}
Due to concurrent transmission of the secondary network, the primary receiver is affected by  the interference from the selected secondary transmitter, 
$S_{k^{\ast}}$.
However, there is transmit power constraint due to primary  QoS  constraint at $R$.
The primary QoS constraint is characterized by  its target outage probability and should be below a threshold level, $\Phi$. The outage probability constraint of the primary network is defined as
\begin{align}\label{desired outage probability}
\mathbb{P}\left[{\Gamma_{TR}}  < \Gamma_0\right]\leq \Phi,
\end{align}
where $\Gamma_{TR}$ is the SINR at $R$, $\Gamma_0 = 2^{\beta}-1$, with $\beta$ as the target rate of the primary network, and $0 < \Phi <1$.
The SINR $\Gamma_{TR}$ is given as  
\begin{align}\label{SINR at P_R}
\Gamma_{TR}= \frac{P_T|h_{TR}|^2}{P_S|h_{S_{k^{\ast}}R}|^2+N_0},
\end{align}
where $P_S|h_{S_{k^{\ast}}R}|^2$ is the interference coming from the secondary small-cell transmitter, $P_S$ is the maximum allowed transmit power of the transmitter which satisfies the primary network 
QoS constraint, $h_{TR}$ is the channel coefficient of the link $T$-$R$, and $h_{S_{k^{\ast}}R}$ is the effective 
channel coefficient of the link $S_{k^{\ast}}$-$R$ due to the particular selection scheme. }

 {We notice from (\ref{desired outage probability}) and (\ref{SINR at P_R}) that $P_S$ can be evaluated from the CDF of $\Gamma_{TR}$ if average power gains of links $h_{S_{k^{\ast}}R}$ and $h_{TR}$ are known.}

\subsection{Interference from the Primary Transmitter}
Due to concurrent transmission, secondary receiver also experience interference from the primary transmitter. In this section, the SINR distributions 
at $D$ and $E$ are derived.
Due to the unreliability of the backhaul, the selected link  may not be active.   
To consider backhaul reliability into the performance analysis, we model it using a Bernoulli random 
variable $\mathbb{I}_k$, for $k=1, \cdots, K$. Including backhaul reliability, the SINR at $D$ can be expressed as 
$\Gamma_{SD}=\mathbb{I}_{k^{\ast}}{\tilde\Gamma}_{SD}$.
$\mathbb{I}_{k^{\ast}}$ represents the backhaul reliability of the selected link, and
\begin{align}
\label{SINR_SD}
{\tilde\Gamma}_{SD}=\frac{ P_S|h_{S_{k^{\ast}}D}|^2}{P_T|h_{TD}|^2+N_0},
\end{align}
 is the SINR if the corresponding backhaul link is active, with $h_{TD}$ as the channel coefficient of the link $T$-$D$ and  
 $h_{S_{k^{\ast}}D}$ as the effective channel coefficient of $S_{k^{\ast}}$-$D$ due to a particular selection scheme.
The distribution of $\Gamma_{SD}$ will be the mixture distribution of $\mathbb{I}_{k^{\ast}}$ and ${\tilde\Gamma}_{SD}$.
As each backhaul link has the same reliability, $\Lambda$, the distribution of $\Gamma_{SD}$ can 
be obtained from the mixture distribution as
\begin{align}
\label{eq_delta}
f_{SD}(x)=(1-\Lambda)\delta(x) + \Lambda {\tilde f}_{SD}, 
\end{align}
where $f_{SD}(x)$ and ${\tilde f}_{SD}(x)$ are the PDFs of $\Gamma_{SD}$ and  ${\tilde \Gamma}_{SD}$, respectively, 
which depend on the particular selection scheme, and $\delta(x)$ is the delta function. 
The CDF of $f_{SD}(x)$ can be obtained by integrating (\ref{eq_delta}), along with finding the 
CDF of ${\tilde\Gamma}_{SD}$.

 {As $E$ can intercept a message only if the selected transmitter transmits, failure of the backhaul leads to failure of both $S_{k^{\ast}}$-$D$ and $S_{k^{\ast}}$-$E$ links.  Hence, the analysis method adopted here includes backhaul reliability in the link $S$-$D$ only. It is not included in the link $S$-$E$, considering that the link $S$-$D$ already includes the backhaul reliability parameter, $\Lambda$, as in (5). Otherwise, in the absence of the $S_{k^{\ast}}$-$D$ link due to backhaul failure, the analysis may include
 $S_{k^{\ast}}$-$E$ according to the independent backhaul link failure assumption.  Conditioned on $S_{k^{\ast}}$ being selected, $E$ always experiences its intercepted signal 
power as independent exponentially distributed.
Hence, while finding the distribution of 
$\Gamma_{SE}$, no backhaul reliability parameter is included.}
As such, SINR at $E$ can be expressed as
\begin{align}\label{SNR_E}
&\Gamma_{SE}=\dfrac{P_S  |h_{S_{k^{\ast}}E}|^2}{P_T  |h_{TE}|^2+N_0}.
\end{align} 
The CDF of $\Gamma_{SE}$ can be derived as
\begin{align}\label{CDF_SNR_E}
&F_{{SE}}(x)=\mathbb{P}\left[\frac{P_S|h_{S_{k^{\ast}}E}|^2}{P_T|h_{TE}|^2+N_0} \leq x \right]
\xd
&=1-\frac{\frac{\lambda_{te}\Gamma_{S}}{\lambda_{se}\Gamma_{T}}}{x+\frac{\lambda_{te}\Gamma_{S}}{\lambda_{se}\Gamma_{T}}} \exp\left(\frac{-\lambda_{se}x}{\Gamma_{S}}\right),
\end{align}
 {where $\Gamma_{T}=\frac{P_T}{N_0}$ and $\Gamma_{S}=\frac{P_S}{N_0}$.}
Then, the PDF can be easily expressed by
\begin{align}\label{PDF_SNR_E}
&f_{{SE}}(x)=\frac{\frac{\lambda_{te}}{\Gamma_{T}}\exp\left(\frac{-\lambda_{se}x}{\Gamma_{S}}\right)}{x+\frac{\lambda_{te}\Gamma_{S}}{\lambda_{se}\Gamma_{T}}}+\frac{\frac{\lambda_{te}\Gamma_{S}}{\lambda_{se}\Gamma_{T}}\exp\left(\frac{-\lambda_{se}x}{\Gamma_{S}}\right)}{\left(x+\frac{\lambda_{te}\Gamma_{S}}{\lambda_{se}\Gamma_{T}}\right)^2}.
\end{align}
\section{Selection Schemes at the Small-Cell Transmitters} \label{section 3}
To improve the secondary secrecy performance, the best transmitter is selected among $K$ small-cell 
transmitters to forward information to $D$.
 {Transmitter selection is carried out before the data transmission begins. These selection schemes, as well as knowledge of backhaul activity, requires channel measurements. 
Acquiring channel knowledge increases energy consumption and complexity of the system. In addition, source selection based on backhaul activity increases extra overhead for the system. To find a trade-off between energy consumption,  system complexity and performance, we consider different source selection schemes using limited available channel knowledge and without knowledge of backhaul activity.}  { The proposed sub-optimal selection schemes, namely sub-optimal transmitter selection (STS), minimal interference selection (MIS), and minimal eavesdropping selection (MES), require knowledge of only one link: either $S_k$-$D$, $S_k$-$R$, or $S_k$-$E$, respectively. On the contrary, the optimal selection (OS) scheme requires knowledge of the $S_k$-$D$, $S_k$-$E$, $S_k$-$R$, $T$-$D$, and $T$-$E$ links. Though, based on these channel knowledge, a small-cell transmitter is selected; there is no guarantee that the transmitter is able to receive any message from the macro-cell BS due to its wireless backhaul failure. }  {The knowledge of backhaul activity  can improve the performance of the proposed selection schemes. However, our results can predict worst case system behavior. How knowledge of backhaul activity  can improve the proposed selection schemes is an interesting future research direction.}
 
 {These selection schemes require channel knowledge at a node capable of running the algorithm for the selection. In the sub-optimal selection methods, the receiver $D$, $R$ or $E$ may take a decision as of which transmitter to be selected using channel estimates from the transmitted pilot symbols. Using a feedback mechanism, the decision may be conveyed to the transmitters at each coherence time, a method generally embedded in many communication protocols to establish communication.  { In the OS scheme, a similar approach can be utilized. Channel estimates can be sent to a central node which then share the decision among small-cell transmitters. The central node can be any node within the network or a dedicated node having access to the channel state information of the network. The degree of available channel knowledge determines whether sub-optimal or optimal algorithms can be implemented. The knowledge of backhaul activity is not assumed at the central node to reduce energy consumption and to make the system simpler.} A detailed discussion on how the channel is acquired and how selection is performed are beyond the scope of this paper. Here, the focus is on the secrecy performance evaluation if any of the proposed selection schemes is in use.}

 {Three sub-optimal selection schemes STS, MIS, and MES, as well as the optimal selection scheme, OS, are proposed in this section. The necessary SINR distributions at $D$, $E$, and $R$ corresponding to the selection schemes, are derived in this section to obtain the secondary non-zero secrecy rate, SOP, and ergodic secrecy rate performances in the subsequent sections. The secondary transmit power limit due to primary outage probability constraint is also evaluated.}

\subsection{Sub-optimal transmitter selection (STS)}
 {Suppose that channel information is available to select the best transmitter to forward the message. Using knowledge of the ${S}_k$-$D$ links, for all $k$, the STS scheme selects the source by finding the maximum power gain among all ${S}_k$-$D$  
links. This can be expressed mathematically as}
\begin{align}\label{OPS_k}
k^* = \arg  \max_{1 \le k \le K} \{ |h_{S_kD}|^2\}.
\end{align}

 {We will now derive the SINR distribution of ${\Gamma}_{SD}$ and secondary transmit power constraint $P_S$.} 
The CDF of ${\tilde\Gamma}_{SD}$ can be evaluated by applying the definition of CDF, with the help 
of (\ref{CDF of h_{S_kD}}) and the PDF of $|h_{TD}|^2$ as
\begin{align}
&{\tilde F}_{{SD}}(x)=\mathbb{P}\left[\frac{ P_S \max_{k=1,...,K}  \{|h_{S_kD}|^2 \}}{P_T|h_{TD}|^2+N_0}\le x\right]
\nonumber\\
&=1-\sum_{k=1}^{K}\binom{K}{k}\frac{(-1)^{k+1}
\frac{\lambda_{td}\Gamma_{S}}{k\lambda_{sd}\Gamma_{T}}}{x+
\frac{\lambda_{td}\Gamma_{S}}{k\lambda_{sd}\Gamma_{T}}}\exp\left( \frac{-k\lambda_{sd}x}{\Gamma_{S}}\right).
 \end{align}
In the previous derivation, the CDF of $ \max\{ |h_{S_kD}|^2\}$, $\forall k$,   is used, which is obtained as
\begin{align}\label{CDF of h_{S_kD}}
F_{{S_{k^{\ast}}D}}(x)&=\mathbb{P}\left[ \max_{1 \le k \le K} \{  |h_{S_kD}|^2\} \le x\right]
= \prod_{k=1}^K \mathbb{P}\left[|h_{S_kD}|^2\le x\right]\nonumber\\
&=1-\sum_{k=1}^{K}\binom{K}{k}(-1)^{k+1}\exp\left(-k\lambda_{sd}x\right).
\end{align} 
The CDF of ${\Gamma}_{SD}$ can be calculated following (\ref{eq_delta}) as 
\begin{align}\label{OTS_CDF_SNR_TD}
 {F}_{SD}(x)=(1-\Lambda)+\Lambda {\tilde F}_{{SD}}(x).
\end{align}

To evaluate $P_S$, the CDF of $\Gamma_{TR}$ needs to be derived for this scheme. 
Irrespective of the particular transmitter selected, $R$ experiences an independent Rayleigh distributed channel; hence,
the CDF of $\Gamma_{TR}$ can be obtained from the definition of CDF and following (\ref{CDF_SNR_E}) as
\begin{align} \label{CDF of Psi}
&F_{{TR}}(x)= 
1-\frac{\frac{\lambda_{sr}\Gamma_T}{\lambda_{tr}\Gamma_{S}}}{x+\frac{\lambda_{sr}\Gamma_T}{\lambda_{tr}\Gamma_{S}}}\textnormal{exp}\left(\frac{-\lambda_{tr}x}{\Gamma_T}\right).
\end{align}
From (\ref{CDF of Psi}) and (\ref{desired outage probability}), the output power of the secondary transmitter can be derived as
\begin{align}\label{P_H}
    P_S=\left\{
                \begin{array}{ll}
                  P_T\lambda_{sr}\xi,\ \ \ \textnormal{if}\ \ \ \xi>0\\
                   0,\ \ \ \ \ \ \ \ \ \ \ \ \textnormal{otherwise},\\
                \end{array}
              \right.
\end{align} 
where
\begin{align} \label{xi}
\xi=\frac{1}{\lambda_{tr}\Gamma_0}\left[\frac{\textnormal{exp}\left(\frac{-\lambda_{tr}\Gamma_0}{\Gamma_{T}}\right)}{1-\Phi}-1\right].
\end{align}

 {Following (\ref{P_H}) and (\ref{xi}) we can observe that for a given primary transmit power and QoS constraint, if the interference from the secondary increases due to better secondary to primary channel condition, secondary transmit power has to be reduced. Further, the secondary transmit power can be increased if the primary channel quality improves.}

\subsection{Minimal Interference Selection (MIS)}
 {In the MIS scheme, the transmitter is selected to guarantee that minimum interference occurs from the selected secondary transmitter to the primary receiver.  The transmitter is selected 
by finding minimum link power gain among $S_k$-$R$ links as }
\begin{align}\label{MIS_k}
k^* = \arg \min_{1 \le k \le K} |h_{S_kR}|^2.
\end{align}
 {This selection scheme requires channel knowledge of all $S_k$-$R$ links.}

 {Now, we will derive the SINR distribution of ${\Gamma}_{SD}$ and secondary transmit power constraint $P_S$ for the MIS scheme.}
Given that the transmitter is selected based on the interference links to $R$, both $D$ and $E$  experience
independent Rayleigh distributed channels. 
Following this argument, the CDF of the SINR distribution, ${\tilde\Gamma}_{SD}$, can be evaluated from the definition of CDF, and with the help of the CDF of $|h_{S_kD}|^2$ and the PDF of $|h_{TD}|^2$ as
\begin{align}
\label{eq_gamma_sd}
&{\tilde F}_{{SD}}(x)=\mathbb{P} \left[\frac{P_S|h_{S_kD}|^2}{P_{T}|h_{TD}|^2+N_0}\le x\right]
\xd
&=1-\frac{\frac{\lambda_{td}\Gamma_{S}}{\lambda_{sd}\Gamma_{T}}}{x+\frac{\lambda_{td}\Gamma_{S}}{\lambda_{sd}\Gamma_{T}}}\exp\left( \frac{-\lambda_{sd}x}{\Gamma_{S}}\right).
\end{align}
The CDF of ${\Gamma}_{SD}$ can then be evaluated using of (\ref{eq_delta}) and (\ref{OTS_CDF_SNR_TD}).

In order to find $P_S$ for this scheme, SINR at $R$ is obtained from (\ref{SINR at P_R}) as  
\begin{align}\label{MIC_SINR at P_RX }
\Gamma_{TR}= \frac{P_T|h_{TR}|^2}{P_S \min_{1 \le k \le K} \{|h_{S_kR}|^2\}+N_0},
\end{align}
the corresponding CDF can be derived as  
\begin{align} \label{MIS_CDF of Psi}
&F_{{TR}}(x)= \mathbb{P}\left[\frac{P_T|h_{TR}|^2}{\min_{1 \le k \le K} P_S|h_{S_kR}|^2+N_0} \leq x \right]
\xd
&=1-\frac{\frac{K\lambda_{sr}\Gamma_T}{\lambda_{tr}\Gamma_{S}}}
{x+\frac{K\lambda_{sr}\Gamma_T}{\lambda_{tr}\Gamma_{S}}}\textnormal{exp}\left(\frac{-\lambda_{tr}x}{\Gamma_T}\right).
\end{align}
As in the STS scheme, $P_S$ can be expressed as in (\ref{P_H}); however, 
$\xi$ is evaluated as follows
\begin{align} \label{MIS_xi}
\xi=\frac{K}{\lambda_{tr}\Gamma_0}\left[\frac{\textnormal{exp}\left(\frac{-\lambda_{tr}\Gamma_0}{\Gamma_{T}}\right)}{1-\Phi}-1\right].
\end{align}
\subsection{Minimal Eavesdropping Selection (MES)}
 {The MES scheme is designed to minimize the eavesdropping from the secondary transmitter. The transmitter is selected by searching the worst link quality among $S_k$-$E$ links, for all $k$. 
This utilizes only channel knowledge of the $h_{S_kE}$ links, as} 
\begin{align}\label{MES_k}
k^* = \arg \min_{1 \le k \le K} |h_{S_kE}|^2.
\end{align}

 {Like other selection schemes, we have to derive the SINR distribution at $D$ and $E$, as well as the secondary transmit power constraint, $P_S$.} 
Given that the transmitter is selected based on the eavesdropping links, $D$ always experiences an independent 
Rayleigh distributed channel, and hence, ${\tilde F}_{{SD}}(x)$ is the same as (\ref{eq_gamma_sd}) 
and subsequently, ${F}_{{SD}}(x)$ can be obtained following (\ref{OTS_CDF_SNR_TD}). 

SINR at $E$ can be obtained according to (\ref{SNR_E}), as
\begin{align}\label{MES_SNR_E}
&\Gamma_{SE}=\dfrac{ P_S  \min_{1 \le k \le K} \{|h_{S_{k}E}|^2\}}{P_T  |h_{TE}|^2+N_0}.
\end{align}
The CDF of the SINR at $E$ can be evaluated from the definition of CDF and following (\ref{PDF_SNR_E}) as 
\begin{align}\label{cdf_mes}
&F_{{SE}}(x)= \mathbb{P}\left[\dfrac{ P_S  \min_{1 \le k \le K} \{|h_{S_{k}E}|^2\}}
{P_T  |h_{TE}|^2+N_0} \leq x \right] \nonumber \\
&= 1- \prod_{k=1}^K \mathbb{P}  \left[ |h_{S_{k}E}|^2 > \frac{\frac{P_T}{N_0}(|h_{TE}|^2+1)x}{\frac{P_S}{N_0}  }\right].
\end{align}
The corresponding PDF can be obtained by differentiating (\ref{cdf_mes}) as
\begin{align}\label{NZ_MES_PDF_SNR_E}
&f_{{SE}}(x)=\frac{\frac{\lambda_{te}}{\Gamma_{T}}
\exp\left(\frac{-K\lambda_{se}x}{\Gamma_{S}}\right)}{x+\frac{\lambda_{te}\Gamma_{S}}
{K\lambda_{se}\Gamma_{T}}}+\frac{\frac{\lambda_{te}\Gamma_{S}}{K\lambda_{se}\Gamma_{T}}
\exp\left(\frac{-K\lambda_{se}x}{\Gamma_{S}}\right)}{\left(x+\frac{\lambda_{te}\Gamma_{S}}
{K\lambda_{se}\Gamma_{T}}\right)^2}.
\end{align}

For this scheme, $P_S$ and $\xi$ are the same as in (\ref{P_H}) and (\ref{xi}), respectively.
 {
\subsection{Optimal Selection (OS)}
As opposed to the sub-optimal selection schemes, where only a single channel information is required, OS requires  all channel information. As all channel information is available, the best transmitter can be selected for which the instantaneous achievable secrecy rate, $C^k_S$, among all $k$, of the secondary network attains maximum. The secrecy rate of the $k^{th}$ link $S_k$-$D$ and $S_k$-$E$ pair is defined as the difference of the main channel achievable rate, $\log_2 (1+ \Gamma_{S_kD})$, and the achievable rate of the wiretap channel,
$\log_2 (1+ \Gamma_{S_kE})$, as \cite{yang2013physical}
\begin{align}
\label{EQ_CAPACITY}
C^k_{S} =\left[\log_2\left(\frac{ 1+ \Gamma_{S_kD}}{1+ \Gamma_{S_kE}}\right)\right]^+ ,
\end{align}
where $[x]^{+}=\max(x,0)$, and $\Gamma_{S_kD}$ and $\Gamma_{S_kE}$ are the SINRs of the $k^{th}$ links  $S_k$-$D$ 
and $S_k$-$E$, respectively. The definitions of these SINRs are the same as in (\ref{SINR_SD}) and (\ref{SNR_E}), respectively, 
for the $k^{th}$ link. 
In OS, the transmitter is selected by searching the maximum $C^{k}_S$ among individual wiretap channels formed by $S_k$-$D$ and $S_k$-$E$ links, for all $k$, as
\begin{align}\label{OS_CS}
k^* = \max_{1 \le k \le K} \{ {C^{k}_S}\}.
\end{align}
Here we note that the optimal transmitter has failure probability of $(1-\Lambda)$ due to backhaul.}

 {For this scheme, $P_S$ and $\xi$ are the same as in (\ref{P_H}) and (\ref{xi}), respectively.}

\section{Non-Zero Secrecy Rate} \label{section 4}
In this section, non-zero secrecy rate is derived for the system with selection schemes. The non-zero secrecy rate is the probability that secrecy capacity is greater than zero \cite{yang2013physical}. 
The probability of non-zero secrecy rate is given as
\begin{align}\label{non_zero_eqt}
\mathbb P [C_{S}>0]
 &
=\mathbb{P}\left[\Gamma_{SD}>\Gamma_{SE}\right]
\xd
&=1-  \int_{0}^{\infty} \CDF{{SD}}{x} \PDF{SE}{x}dx.  
\end{align}
Substituting $\CDF{{SD}}{x}$ and $\PDF{SE}{x}$ for a particular selection scheme, which we already derived in the previous section, 
the corresponding non-zero secrecy rate can be obtained as in the following subsections. 

 {Due to the assumption of independent and identically distributed channel conditions of $S_k$-$D$, $S_k$-$E$, and $S_k$-$R$ links for all $k$, respectively, $F_{SD}(x)$ and $f_{SE}(x)$ already incorporate information related to the particular selection.  Hence, non-zero secrecy rate can be obtained by  directly evaluating (\ref{non_zero_eqt}).  
In the independent non-identically distributed case, non-zero secrecy rate can be evaluated by finding the probability that the $k^{th}$ transmitter is selected, and then, calculating the non-zero secrecy rate for that selection using (\ref{non_zero_eqt}). Finally, with the help of  the law of total probability, the non-zero secrecy rate can be evaluated. Similarly, SOP and ergodic secrecy rate can also be evaluated. The derivation method explained can be found in \cite{Kundu_relsel}. } 

\subsection{sub-optimal transmitter selection (STS)}
The non-zero secrecy rate of the STS scheme can be obtained in the following corollary. 
\begin{corollary}
\label{corrolar1}
The non-zero secrecy rate of the STS scheme is 
\begin{align}
\label{NZ_rate OTS}
\mathbb P(C_{S}>0)&=\sum_{k=1}^{K}\binom{K}{k}\Lambda(-1)^{k+1}\frac{\lambda_{te}\lambda_{td}\Gamma_{S}}{k\lambda_{sd}\Gamma_{T}^2} I_1
\xd
&+\sum_{k=1}^{K}\binom{K}{k}\Lambda(-1)^{k+1}\frac{\lambda_{te}\lambda_{td}\Gamma_{S}^2}{k\lambda_{se}\lambda_{sd}\Gamma_{T}^2} I_2,
\end{align} 
where 
\begin{align}
\label{eq_nonzero_ots_i1}
I_1=&\frac{\exp\left(ac\right)\Ei\left(-ac\right)}{a-b}
-\frac{\exp\left(bc\right)\Ei\left(-bc\right)}{a-b},
\end{align}
\begin{align}
\label{eq_nonzero_ots_i2}
I_2=&-\frac{\exp\left(ac\right)\Ei\left(-ac\right)}{\left(a-b\right)^2}
+\frac{\exp\left(bc\right)\Ei\left(-bc\right)}{\left(a-b\right)^2}
\xd
+&\frac{\left(c\exp\left(bc\right)\Ei\left(-bc\right)+\frac{1}{b}\right)}{a-b}, 
\end{align}
with $a=\frac{\lambda_{td}\Gamma_{S}}{k\lambda_{sd}\Gamma_{T}}$, 
$b=\frac{\lambda_{te}\Gamma_S}{\lambda_{se}\Gamma_T}$, 
$c=\frac{k\lambda_{sd}+\lambda_{se}}{\Gamma_S}$, and $\Ei(x)$ is the exponential integral.

\begin{proof}
Non-zero secrecy rate in (\ref{NZ_rate OTS}) can  be obtained by evaluating (\ref{non_zero_eqt}) with the help of 
(\ref{PDF_SNR_E}) and (\ref{OTS_CDF_SNR_TD}) where we define 
\begin{align}\label{NZ I_1}
I_1=&\int_{0}^{\infty}\frac{1}{(x+a)(x+b)}\exp\left( -cx\right)
dx,
\end{align}
\begin{align}\label{NZ I_2}
I_2=&\int_{0}^{\infty}\frac{1}{\left(x+a\right)\left(x+b\right)^2}\exp\left( -cx\right)
dx.
\end{align}
By utilizing partial fraction to transform multiplication into summations, the integrals in (\ref{NZ I_1}) and (\ref{NZ I_2}) 
can be evaluated easily by following   
\begin{align}\label{PF_NZ_OTS_I_1}
\frac{1}{(x+a)(x+b)}=-\frac{1}{(a-b)(x+a)}+\frac{1}{(a-b)(x+b)},
\end{align}
\begin{align}\label{PF_NZ_OTS_I_2}
\frac{1}{(x+a)(x+b)^2}&=\frac{1}{(a-b)^2(x+a)}-\frac{1}{(a-b)^2(x+b)}
\xd
&+\frac{1}{(a-b)(x+b)^2}.
\end{align}
For the final result, integral solutions of the form \cite{jeffrey2007table}, eq. (3.352.4)
and \cite{jeffrey2007table}, eq. (3.353.3) are considered to get (\ref{eq_nonzero_ots_i1}) 
and (\ref{eq_nonzero_ots_i2}), respectively.
\end{proof}
\end{corollary}
\subsection{Minimal Interference Selection (MIS)}
Substituting $\CDF{{SD}}{x}$ and $\PDF{SE}{x}$ corresponding to this selection scheme 
into (\ref{non_zero_eqt}), and following Corollary \ref{corrolar1},
the non-zero secrecy rate can be evaluated as
\begin{align}
\mathbb P(C_{S}>0)&=\frac{\Lambda\lambda_{te}\lambda_{td}\Gamma_{S}}{\lambda_{sd}\Gamma_{T}^2} I_1
+\frac{\Lambda\lambda_{te}\lambda_{td}\Gamma_{S}^2}{\lambda_{se}\lambda_{sd}\Gamma_{T}^2} I_2,
\end{align}
where $I_1$ and $I_2$ are the same as (\ref{NZ I_1}) and (\ref{NZ I_2}), 
respectively, with $a=\frac{\lambda_{td}\Gamma_{S}}{\lambda_{sd}\Gamma_{T}}$, 
$b=\frac{\lambda_{te}\Gamma_S}{\lambda_{se}\Gamma_T}$, and 
$c=\frac{\lambda_{sd}+\lambda_{se}}{\Gamma_S}$.
\subsection{Minimal Eavesdropping Selection (MES)}
Similar to the previous subsections, substituting  $\CDF{{SD}}{x}$ and $\PDF{SE}{x}$ 
corresponding to this selection scheme into (\ref{non_zero_eqt}), the non-zero secrecy rate can be evaluated as
\begin{align}
\mathbb P(C_{S}>0)&=\frac{\Lambda\lambda_{te}\lambda_{td}\Gamma_{S}}{\lambda_{sd}\Gamma_{T}^2} I_1
+\frac{\Lambda\lambda_{te}\lambda_{td}\Gamma_{S}^2}{K\lambda_{se}\lambda_{sd}\Gamma_{T}^2} I_2,
\end{align}
where $I_1$ and $I_2$ are same as (\ref{NZ I_1}) and (\ref{NZ I_2}), respectively, with 
$a=\frac{\lambda_{td}\Gamma_{S}}{\lambda_{sd}\Gamma_{T}}$, $b=\frac{\lambda_{te}
\Gamma_S}{K\lambda_{se}\Gamma_T}$, and $c=\frac{\lambda_{sd}+K\lambda_{se}}{\Gamma_S}$.
 {
\subsection{Optimal Selection (OS)}
The probability that the secrecy rate is greater than zero for the OS can be obtained by finding the probability 
that the maximum achievable $C^k_S$, among all $k$, is higher than zero. In this case we notice that the channel coefficients $h_{TD}$ and $h_{TE}$ are common to the $C^k_S$ expression for all $k$; in other words, $C^k_S$ for all $k$ are not independent. As a result, to find the non-zero secrecy rate we first find the probability for a given channel power gain, $|h_{TD}|^2$ and $|h_{TE}|^2$, and then average it over all possibilities.} 

 {Towards deriving the non-zero secrecy of the scheme, we need to find the conditional distribution of $\Gamma_{S_kD}$ and $\Gamma_{S_kE}$ conditioned on $|h_{TD}|^2$ and $|h_{TE}|^2$, respectively. These can be easily obtained from (\ref{SINR_SD}) using the definition of CDF and the corresponding PDF. 
The conditional CDF of $\Gamma_{S_kD}$ conditioned on $|h_{TD}|^2$ can be derived as
\begin{align}
\label{OS_CDF}
&F_{{S_kD}||h_{TD}|^2}(x|y)
=\left[1-\exp{\left(\frac{-x\lambda_{sd}\left(y\Gamma_T+1\right)}{\Gamma_S}\right)}\right].
\end{align}	
Further, the corresponding PDF can be obtained as 
\begin{align}
\label{OS_PDF}
&f_{{S_kD}||h_{TD}|^2}(x|y)
=\frac{\lambda_{sd}\left(y\Gamma_T+1\right)}{\Gamma_S}\exp\left(\frac{-x\lambda_{sd}\left(y\Gamma_T+1\right)}{\Gamma_S}\right).
\end{align}
Similarly, CDF and PDF of $\Gamma_{S_kE}$ conditioned on $|h_{TE}|^2$ can be obtained from the above equations by replacing $D$ with $E$.  } 

 {With the help of the probability that the maximum achievable $C^k_S$ among all $k$ is higher than zero, the non-zero secrecy rate of the OS scheme including backhaul reliability can be evaluated as 
 \begin{align}
\label{OS_NZ_K}
\mathbb P(C_{S}>0)=(1-\Lambda)\times 0 + \Lambda \mathbb{P} [ \max_{1 \le k \le K}\{C^k_{S}\}>0],
\end{align}
 where 
\begin{align}\label{non_zero_eqt_OS}
&\mathbb P [ \max_{1 \le k \le K} \{C^k_{S}\}>0]=1-\mathbb P [ \max_{1 \le k \le K} \{C^k_{S}\} \le 0]
\xd
&=1-\mathbb E_{|h_{TD}|^2 }\mathbb E_{|h_{TE}|^2}\left[\mathbb P [C^k_{S}\le 0||h_{TD}|^2, |h_{TE}|^2]\right]^K
\xd
&=1-\int_{0}^{\infty}\int_{0}^{\infty}\left[\frac{\lambda_{sd}
\left(\Gamma_Ty+1\right)}{\lambda_{sd}\left(\Gamma_Ty+1\right)+\lambda_{se}\left(\Gamma_Tx+1\right)}\right]^K
\xd
&\times\lambda_{te}\exp\left(-\lambda_{te}x\right)\lambda_{td}\exp\left(-\lambda_{td}y\right)dxdy.
\end{align} 
The above derivation uses (\ref{EQ_CAPACITY}), (\ref{OS_CDF}), and (\ref{OS_PDF}).
A closed-form expression is very difficult to achieve in the above derivation; hence, we evaluate it with the help of the Mathematica software.}
\section{Secrecy Outage Probability} \label{section 5}
The SOP is defined as the probability that the secrecy rate is less than a certain threshold secrecy rate, $R_{th}$. 
The SOP can be expressed as \cite{wang2015security,yang2013transmit}
\begin{align}\label{secrecy OP equation}
\mathcal{P}_{out} (R_{th})
&= \mathbb P\left[C_{S} < R_{th}\right],\nonumber \\
&= \int_{0}^{\infty} F_{SD}(\rho(x+1)-1) f_{SE}(x) dx,
\end{align}
where $\rho = 2^{R_{th}}-1$.
In the following subsections, SOP is derived for the proposed selection schemes.
\subsection{sub-optimal transmitter selection (STS)}
By substituting $F_{SD}$ and $f_{SE}$ from (\ref{OTS_CDF_SNR_TD}) and (\ref{PDF_SNR_E}) into (\ref{secrecy OP equation}),
and following the derivation method as in Corollary \ref{corrolar1}, the SOP can be evaluated as
\begin{align}
\mathcal{P}_{out} (R_{th})=1-A I_1-B I_2,
\end{align}
where $I_1$ and $I_2$ are the same as (\ref{NZ I_1}) and (\ref{NZ I_2}), respectively, with  
$a=\frac{\lambda_{td}\Gamma_{S}+k\rho\lambda_{sd}\Gamma_{T}-k\lambda_{sd}\Gamma_{T}}
{k\rho\lambda_{sd}\Gamma_{T}}$, $b=\frac{\lambda_{te}\Gamma_S}{\lambda_{se}\Gamma_T}$, 
$c=\frac{k\rho\lambda_{sd}+\lambda_{se}}{\Gamma_S}$, 
and 
\begin{align}
A=&\sum_{k=1}^{K}\binom{K}{k}\Lambda(-1)^{k+1}\frac{\lambda_{te}\lambda_{td}\Gamma_{S}}{k\rho\lambda_{sd}\Gamma_{T}^2}\exp\left( \frac{-k\lambda_{sd}\left(\rho-1\right)}{\Gamma_{S}}\right),
\\
B=&\sum_{k=1}^{K}\binom{K}{k}\Lambda(-1)^{k+1}\frac{\lambda_{te}\lambda_{td}\Gamma_{S}^2}{k\rho\lambda_{se}\lambda_{sd}\Gamma_{T}^2}\exp\left( \frac{-k\lambda_{sd}\left(\rho-1\right)}{\Gamma_{S}}\right).
\end{align}
\subsection{Minimal Interference Selection (MIS)}
The SOP can be evaluated similarly as
\begin{align}
\mathcal{P}_{out} (R_{th})&=1-\frac{\Lambda\lambda_{te}\lambda_{td}\Gamma_{S}}{\rho\lambda_{sd}\Gamma_{T}^2}\exp\left( \frac{-\lambda_{sd}\left(\rho-1\right)}{\Gamma_{S}}\right) I_1
\xd
&-\frac{\Lambda\lambda_{te}\lambda_{td}\Gamma_{S}^2}{\rho\lambda_{se}\lambda_{sd}\Gamma_{T}^2}\exp\left( \frac{-\lambda_{sd}\left(\rho-1\right)}{\Gamma_{S}}\right) I_2,
\end{align}
where $I_1$ and $I_2$ are the same as (\ref{NZ I_1}) and (\ref{NZ I_2}), respectively, with 
$a=\frac{\lambda_{td}\Gamma_{S}+\rho\lambda_{sd}\Gamma_{T}
-\lambda_{sd}\Gamma_{T}}{\rho\lambda_{sd}\Gamma_{T}}$,
$b=\frac{\lambda_{te}\Gamma_S}{\lambda_{se}\Gamma_T}$, and 
$c=\frac{\rho\lambda_{sd}+\lambda_{se}}{\Gamma_S}$.
\subsection{Minimal Eavesdropping Selection (MES)}
Similar to the previous two subsections, the SOP can be evaluated as
\begin{align}
\mathcal{P}_{out} (R_{th})&=1-\frac{\Lambda\lambda_{te}\lambda_{td}\Gamma_{S}}{\rho\lambda_{sd}\Gamma_{T}^2}\exp\left( \frac{-\lambda_{sd}\left(\rho-1\right)}{\Gamma_{S}}\right) I_1
\xd
&-\frac{\Lambda\lambda_{te}\lambda_{td}\Gamma_{S}^2}{\rho K\lambda_{se}\lambda_{sd}\Gamma_{T}^2}\exp\left( \frac{-\lambda_{sd}\left(\rho-1\right)}{\Gamma_{S}}\right) I_2,
\end{align}
where $I_1$ and $I_2$ are the same as (\ref{NZ I_1}) and (\ref{NZ I_2}) respectively, with 
$a=\frac{\lambda_{td}\Gamma_{S}+\rho\lambda_{sd}\Gamma_{T}-\lambda_{sd}\Gamma_{T}}
{\rho\lambda_{sd}\Gamma_{T}}$, $b=\frac{\lambda_{te}\Gamma_S}
{K\lambda_{se}\Gamma_T}$, and $c=\frac{\lambda_{sd}+K\lambda_{se}}
{\Gamma_S}$. 
 {
\subsection{Optimal Selection (OS)}
As in non-zero secrecy rate, SOP of the OS scheme can be evaluated including the backhaul reliability by finding the probability that the maximum $C^k_S$, among all $k$, is less than the required threshold secrecy rate, as
\begin{align}
\label{OS_SOP}
\mathcal{P}_{out} (R_{th})=(1-\Lambda)+ \Lambda \mathbb P\left[ \max_{1 \le k \le K} \{C^k_{S}\} < R_{th}\right],
\end{align}
where 
\begin{align}\label{secrecy OP equation OS}
&\mathbb P\left[ \max_{1 \le k \le K}\{C_{S}\} < R_{th}\right]\xd
&=\mathbb E_{|h_{TD}|^2 }\mathbb E_{|h_{TE}|^2}\left[\mathbb P [C^k_{S}\le R_{th}||h_{TD}|^2, |h_{TE}|^2]\right]^K
\xd
&=\int_0^\infty\int_0^\infty 
\bigg[1-\frac{\lambda_{se}\left(\Gamma_Ty+1\right)}{\rho\lambda_{sd}\left(\Gamma_Tx+1\right)+\lambda_{se}\left(\Gamma_Ty+1\right)}\xd
&\exp\left(\frac{-\lambda_{sd}\left(\rho-1\right)\left(\Gamma_T x+1\right)}{\Gamma_S}\right)
\bigg]^K \xd
&\times \lambda_{td}\exp\left(-\lambda_{td}x\right)\lambda_{te}\exp\left(-\lambda_{te}y\right)dxdy.
\end{align}
The solution expressed above follows the method previously applied to derive the non-zero secrecy rate of the OS scheme 
with the help of (\ref{EQ_CAPACITY}), (\ref{OS_CDF}), and (\ref{OS_PDF}). The above equation is computed with the help of the Mathematica software, as it is difficult to achieve the closed-form solution.
}

\section{Ergodic Secrecy Rate} \label{section 6}
The ergodic secrecy rate is defined as the  secrecy rate averaged over all channel realizations, being given by \cite{hoang2015cooperative} 
\begin{align}  \label{erg_eqt}
\mathcal{C}_{erg}
&= \dfrac{1}{\ln(2)}\int_{0}^{\infty} \dfrac{\CDF{SE}{x}}{1+x}[1-\CDF{{SD}}{x}] dx.
\end{align}
The following subsections derive the expressions for the ergodic secrecy rate by substituting 
$F_{SE}$(x) and $F_{SD}(x)$ already derived in  Sections \ref{section 2} and \ref{section 3}, respectively, 
into (\ref{erg_eqt}).  

\subsection{sub-optimal transmitter selection (STS)}

\begin{corollary}
\label{corrolary2}
The ergodic secrecy capacity of the STS scheme is given as
\begin{align}
\label{eq_ergodic_ots1}
\mathcal{C}_{erg}=\dfrac{1}{\ln(2)}\sum_{k=1}^{K}\binom{K}{k}\Lambda(-1)^{k+1}\left(I_1+I_2\right),
\end{align}
where $I_1$ and $I_2$ are given respectively as
\begin{align}
\label{eq_ergodic_ots2}
I_1=&-b\Bigg(\frac{\exp\left(c_1\right)\Ei\left(-c_1\right)}{b-1}
+\frac{\exp\left(bc_1\right)\Ei\left(-bc_1\right)}{1-b}\Bigg),
\end{align}
\begin{align}
\label{eq_ergodic_ots3}
I_2=&ab\Bigg(\frac{\exp\left(c_2\right)\Ei\left(-c_2\right)}{\left(a-1\right)\left(b-1\right)}+\frac{\exp\left(ac_2\right)\Ei\left(-ac_2\right)}{\left(a-1\right)\left(a-b\right)}
\xd
+&\frac{\exp\left(bc_2\right)\Ei\left(-bc_2\right)}{\left(b-1\right)\left(b-a\right)}\Bigg),
\end{align}
with  $a=\frac{\lambda_{te}\Gamma_{S}}{\lambda_{se}\Gamma_{T}}$, 
$b=\frac{\lambda_{td}\Gamma_S}{k\lambda_{sd}\Gamma_T}$, 
$c_1=\frac{k\lambda_{sd}}{\Gamma_S}$, and $c_2=\frac{k\lambda_{sd}+\lambda_{se}}{\Gamma_S}$.

\begin {proof}
To get the ergodic secrecy capacity of the STS scheme, we substitute (\ref{CDF_SNR_E}) and (\ref{OTS_CDF_SNR_TD}) into (\ref{erg_eqt}) to get the result in (\ref{eq_ergodic_ots1}) where 
\begin{align}
\label{erg I_1}
I_1=&\int_{0}^{\infty} \frac{b}{\left(x+1\right)\left(x+b\right)}\exp\left(-c_1 x\right)dx, 
\end{align}
\begin{align}
\label{erg I_2}
I_2=&-\int_{0}^{\infty}\frac{ab}{\left(x+1\right)\left(x+a\right)\left(x+b\right)}\exp\left(-c_2 x\right)dx.
\end{align}
To solve $I_1$ and $I_2$, partial fraction is again used as follows
\begin{align}\label{PF_ERG_OTS_I_1}
\frac{1}{(x+1)(x+b)}=\frac{1}{(b-1)(x+1)}+\frac{1}{(1-b)(x+b)},
\end{align}
\begin{align}\label{PF_ERG_OTS_I_2}
&\frac{1}{(x+1)(x+a)(x+b)}=-\frac{1}{(a-1)(b-1)(x+1)}
\xd
&-\frac{1}{(a-1)(a-b)(x+a)}-\frac{1}{(b-1)(b-a)(x+b)}.
\end{align}
For the final solution in (\ref{erg I_1}) and (\ref{erg I_2}), 
integral solution of the form \cite{jeffrey2007table}, eq. (3.352.4) is considered 
to get (\ref{eq_ergodic_ots2}) and (\ref{eq_ergodic_ots3}), respectively.

\end {proof}
\end{corollary}

\subsection{Minimal Interference Selection (MIS)}
Following Corollary \ref{corrolary2}, the ergodic secrecy capacity of this scheme is given as
\begin{align}
\mathcal{C}_{erg}=\Lambda\left(I_1+I_2\right),
\end{align}
where $I_1$ and $I_2$ are same as (\ref{erg I_1}) and (\ref{erg I_2}) respectively, with 
$a=\frac{\lambda_{te}\Gamma_{S}}{\lambda_{se}\Gamma_{T}}$, $b=\frac{\lambda_{td}\Gamma_S}{\lambda_{sd}\Gamma_T}$, 
 $c_1=\frac{\lambda_{sd}}{\Gamma_S}$, and $c_2=\frac{\lambda_{sd}+\lambda_{se}}{\Gamma_S}$. 
\subsection{Minimal Eavesdropping Selection (MES)}
Following Corollary \ref{corrolary2}, the ergodic secrecy capacity of the scheme can be evaluated as
\begin{align}
\mathcal{C}_{erg}=\Lambda\left(I_1+I_2\right),
\end{align}
where $I_1$ and $I_2$ 
are the same as (\ref{erg I_1}) and (\ref{erg I_2}), respectively, with $a=\frac{\lambda_{te}\Gamma_{S}}{K\lambda_{se}\Gamma_{T}}$, 
$b=\frac{\lambda_{td}\Gamma_S}{\lambda_{sd}\Gamma_T}$, 
$c_1=\frac{\lambda_{sd}}{\Gamma_S}$, and 
$c_2=\frac{\lambda_{sd}+K\lambda_{se}}{\Gamma_S}$. 
\section{Asymptotic Secrecy Analysis}\label{section 7}
To provide full insights into the impacts of unreliable backhaul connections, 
the asymptotic analyses of non-zero secrecy rate, SOP, and ergodic secrecy capacity 
of the three selection schemes are carried out by assuming $\gamma_{T} \rightarrow \infty$.
\subsection{Non-Zero Secrecy Rate}
\subsubsection{sub-optimal transmitter selection (STS)}
\begin{corollary}\label{ASY_NZ_COR}
The asymptotic non-zero secrecy rate for the STS scheme is given by 
\begin{align}
\label{corollary3}
Pr(C_{S}>0)\approx&\frac{ln\left(b\right)A\lambda_{te}\lambda_{sr}\xi}{\lambda_{se}\left(a-b\right)^2}
+\frac{A\lambda_{te}\lambda_{sr}\xi}{\lambda_{se}b\left(a-b\right)}
\xd
-&\frac{ln\left(a\right)A\lambda_{te}\lambda_{sr}\xi}{\lambda_{se}\left(a-b\right)^2},
\end{align}
where $\xi\approx\frac{1}{\lambda_{tr}\Gamma_0}\left[\frac{\Phi}{1-\Phi}\right]$.
\begin{proof}
At high SINR, by assuming $\textnormal{exp}\left(\frac{-\lambda_{tr}\Gamma_0}{\Gamma_{T}}\right)$ tends to unity and 
$\sum_{k=1}^{K}\binom{K}{k}\Lambda(-1)^{k+1}
\frac{\lambda_{te}\lambda_{td}\Gamma_{S}}{k\lambda_{sd}\Gamma_{T}^2}  I_1$ tends to zero as $\gamma_{T} \rightarrow \infty$, 
(\ref{NZ_rate OTS}) can be approximated to
\begin{align}\label{ASY_NZ_OTS}
Pr(C_{S}>0)\approx& A I_2,
\end{align}
where 
\begin{align}
A=\sum_{k=1}^{K}\binom{K}{k}\Lambda(-1)^{k+1}\frac{\lambda_{sr}\lambda_{td}\xi}{k \lambda_{sd}}
\end{align}
and 
\begin{align} \label{ASY_OTS_xi}
\xi\approx\frac{1}{\lambda_{tr}\Gamma_0}\left[\frac{\Phi}{1-\Phi}\right].
\end{align}
Further, by assuming $\exp\left( -cx\right)$ tending to unity as $c$ tends to zero due to $\Gamma_{S} \rightarrow \infty$ in 
(\ref{eq_nonzero_ots_i1}) and (\ref{eq_nonzero_ots_i2}), $I_2$ can be simplified to 
\begin{align}\label{ASY_NZ_OTS_I_2}
I_2\approx& \int_{0}^{\infty}\frac{1}{\left(x+a\right)\left(x+b\right)^2}dx,
\end{align}
where $a=\frac{\lambda_{sr}\lambda_{td}\xi}{k \lambda_{sd}}$, 
$b=\frac{\lambda_{te}\lambda_{sr}\xi}{\lambda_{se}}$.
Substituting (\ref{ASY_NZ_OTS_I_2}) into (\ref{ASY_NZ_OTS}), an asymptotic non-zero secrecy rate can be derived as
\begin{align}\label{ASY_NZ_APPROX}
Pr(C_{S}>0)\approx \int_{0}^{\infty}\frac{A}{\left(x+a\right)\left(x+b\right)^2}dx.
\end{align}
Finally, the partial fraction method of (\ref{PF_NZ_OTS_I_2}) is implemented in (\ref{ASY_NZ_APPROX}) to obtain the desired result
in (\ref{corollary3}). 
 
\end{proof}

\end{corollary}

\subsubsection{Minimal Interference Selection (MIS)}
Following Corollary \ref{ASY_NZ_COR}, the asymptotic non-zero secrecy rate expression of this scheme can be derived as
\begin{align}
Pr(C_{S}>0)\approx&\frac{ln\left(b\right)A\lambda_{te}\lambda_{sr}\xi}{\lambda_{se}\left(a-b\right)^2}
+\frac{A\lambda_{te}\lambda_{sr}\xi}{\lambda_{se}b\left(a-b\right)}
\xd
-&\frac{ln\left(a\right)A\lambda_{te}\lambda_{sr}\xi}{\lambda_{se}\left(a-b\right)^2},
\end{align}
where $A=\frac{\Lambda\lambda_{sr}\lambda_{td}\xi}{ \lambda_{sd}}$, 
$a=\frac{\lambda_{sr}\lambda_{td}\xi}{ \lambda_{sd}}$, $b=\frac{\lambda_{te}\lambda_{sr}\xi}{\lambda_{se}}$, 
and 
\begin{align} \label{ASY_MIS_xi}
\xi&\approx\frac{K}{\lambda_{tr}\Gamma_0}\left[\frac{\Phi}{1-\Phi}\right].
\end{align}

\subsubsection{Minimal Eavesdropping Selection (MES)}
Following Corollary \ref{ASY_NZ_COR}, the asymptotic non-zero secrecy rate expression of this scheme can be derived as
\begin{align}
Pr(C_{S}>0)\approx&\frac{ln\left(b\right)A\lambda_{te}\lambda_{sr}\xi}{K\lambda_{se}\left(a-b\right)^2}
+\frac{A\lambda_{te}\lambda_{sr}\xi}{K\lambda_{se}b\left(a-b\right)}
\xd
-&\frac{ln\left(a\right)A\lambda_{te}\lambda_{sr}\xi}{K\lambda_{se}\left(a-b\right)^2},
\end{align}
where $A=\frac{\Lambda\lambda_{sr}\lambda_{td}\xi}{ \lambda_{sd}}$,
$a=\frac{\lambda_{sr}\lambda_{td}\xi}{ \lambda_{sd}}$, and $b=\frac{\lambda_{te}\lambda_{sr}\xi}{K\lambda_{se}}$, and 
$\xi$ is the same as in (\ref{ASY_OTS_xi}).

\subsubsection{Optimal Selection (OS)}
Following Corollary \ref{ASY_NZ_COR}, the asymptotic non-zero secrecy rate expression of this scheme can be derived as
\begin{align}
Pr(C_{S}>0)\approx&1-\Big[1-\frac{ln\left(b\right)A\lambda_{te}\lambda_{sr}\xi}{\lambda_{se}\left(a-b\right)^2}
-\frac{A\lambda_{te}\lambda_{sr}\xi}{\lambda_{se}b\left(a-b\right)}
\xd
+&\frac{ln\left(a\right)A\lambda_{te}\lambda_{sr}\xi}{\lambda_{se}\left(a-b\right)^2}\Big]^K,
\end{align}
where $A=\frac{\Lambda\lambda_{sr}\lambda_{td}\xi}{ \lambda_{sd}}$, 
$a=\frac{\lambda_{sr}\lambda_{td}\xi}{ \lambda_{sd}}$, $b=\frac{\lambda_{te}\lambda_{sr}\xi}{\lambda_{se}}$, 
and $\xi$ is the same as in (\ref{ASY_OTS_xi}).
 
\subsection{Secrecy Outage Probability}
\subsubsection{sub-optimal transmitter selection (STS)}
Following Corollary \ref{ASY_NZ_COR}, the asymptotic SOP expression for this scheme can be derived as
\begin{align}
\mathcal{P}_{out} (\rho)\approx&1-\frac{ln\left(b\right)A\lambda_{te}\lambda_{sr}\xi}{\lambda_{se}\left(a-b\right)^2}
-\frac{A\lambda_{te}\lambda_{sr}\xi}{\lambda_{se}b\left(a-b\right)}
\xd
+&\frac{ln\left(a\right)A\lambda_{te}\lambda_{sr}\xi}{\lambda_{se}\left(a-b\right)^2},
\end{align}
where \begin{align}
A=\sum_{k=1}^{K}\binom{K}{k}\Lambda(-1)^{k+1}\frac{\lambda_{sr}\lambda_{td}\xi}{k \rho\lambda_{sd}},
\end{align}
$a=\frac{k(\rho-1)\lambda_{sd}+\lambda_{sr}\lambda_{td}\xi}{\rho k\lambda_{sd}}$,  
$b=\frac{\lambda_{te}\lambda_{sr}\xi}{\lambda_{se}}$, and $\xi$ is the one corresponding to the STS scheme 
as in (\ref{ASY_OTS_xi}). 
\subsubsection{Minimal Interference Selection (MIS)}
Following Corollary \ref{ASY_NZ_COR}, the asymptotic SOP expression for this scheme can be derived as
\begin{align}
\mathcal{P}_{out} (\rho)\approx&1-\frac{ln\left(b\right)A\lambda_{te}\lambda_{sr}\xi}{\lambda_{se}\left(a-b\right)^2}
-\frac{A\lambda_{te}\lambda_{sr}\xi}{\lambda_{se}b\left(a-b\right)}
\xd
+&\frac{ln\left(a\right)A\lambda_{te}\lambda_{sr}\xi}{\lambda_{se}\left(a-b\right)^2},
\end{align}
where $a=\frac{(\rho-1)\lambda_{sd}+\lambda_{sr}\lambda_{td}\xi}{\rho\lambda_{sd}}$, 
$b=\frac{\lambda_{te}\lambda_{sr}\xi}{\lambda_{se}}$, $A=\frac{\Lambda\lambda_{sr}\lambda_{td}\xi}{\rho\lambda_{sd}}$, and 
$\xi$ is that corresponding to the scheme as in (\ref{ASY_MIS_xi}).
\subsubsection{Minimal Eavesdropping Selection (MES)}
Similarly, following Corollary \ref{ASY_NZ_COR}, 
the asymptotic SOP expression for this scheme can be derived as
\begin{align}
\mathcal{P}_{out} (\rho)\approx&1-\frac{ln\left(b\right)A\lambda_{te}\lambda_{sr}\xi}{K\lambda_{se}\left(a-b\right)^2}
-\frac{A\lambda_{te}\lambda_{sr}\xi}{K\lambda_{se}b\left(a-b\right)}
\xd
+&\frac{ln\left(a\right)A\lambda_{te}\lambda_{sr}\xi}{K\lambda_{se}\left(a-b\right)^2},
\end{align}
where $A=\frac{\Lambda\lambda_{sr}\lambda_{td}\xi}{\rho\lambda_{sd}}$, 
$a=\frac{(\rho-1)\lambda_{sd}+\lambda_{sr}\lambda_{td}\xi}{\rho\lambda_{sd}}$, 
$b=\frac{\lambda_{te}\lambda_{sr}\xi}{K\lambda_{se}}$, and $\xi$ follows (\ref{ASY_OTS_xi}).

\subsubsection{Optimal Selection (OS)}
Following Corollary \ref{ASY_NZ_COR}, the asymptotic SOP expression for this scheme can be derived as
\begin{align}
\mathcal{P}_{out} (\rho)\approx&\Big[1-\frac{ln\left(b\right)A\lambda_{te}\lambda_{sr}\xi}{\lambda_{se}\left(a-b\right)^2}
-\frac{A\lambda_{te}\lambda_{sr}\xi}{\lambda_{se}b\left(a-b\right)}
\xd
+&\frac{ln\left(a\right)A\lambda_{te}\lambda_{sr}\xi}{\lambda_{se}\left(a-b\right)^2}\Big]^K,
\end{align}
where $a=\frac{(\rho-1)\lambda_{sd}+\lambda_{sr}\lambda_{td}\xi}{\rho\lambda_{sd}}$, 
$b=\frac{\lambda_{te}\lambda_{sr}\xi}{\lambda_{se}}$, $A=\frac{\Lambda\lambda_{sr}\lambda_{td}\xi}{\rho\lambda_{sd}}$, and 
$\xi$ is that corresponding to the scheme as in (\ref{ASY_OTS_xi}). 
\subsection{Ergodic Secrecy Capacity}
\subsubsection{sub-optimal transmitter selection (STS)}
\begin{corollary}
\label{ASY_ERG_COR}
The asymptotic ergodic secrecy capacity for the STS scheme is given by 
\begin{align}\label{eq_asym_erg_ots}
\mathcal{C}_{erg}
=&A ab\left(\frac{ln(a)}{(a-1)(a-b)}+\frac{ln(b)}{(b-1)(b-a)}\right)
\xd
-&A b\frac{ln(b)}{1-b},
\end{align}
where 
\begin{align}
A=\frac{1}{ln(2)}\sum_{k=1}^{K}\binom{K}{k}\Lambda(-1)^{k+1}.
\end{align}
\begin{proof}
Following a similar approach as in the previous subsections, by approximating 
$\exp\left(-c_1 x\right)$ and $\exp\left(-c_2 x\right)$ tending to unity as $c_1$ and $c_2$ tend to zero at high SINR,
the asymptotic ergodic secrecy capacity can be derived as
\begin{align}
\mathcal{C}_{erg}=A\left(I_1+I_2\right),
\end{align}
where
\begin{align}
\label{ASY_erg I_1}
I_1\approx\int_{0}^{\infty} \frac{b}{\left(x+1\right)\left(x+b\right)}dx,\\
I_2\approx-\int_{0}^{\infty}\frac{ab}{\left(x+1\right)\left(x+a\right)\left(x+b\right)}dx,
\end{align}
\begin{align}
A=\frac{1}{ln(2)}\sum_{k=1}^{K}\binom{K}{k}\Lambda(-1)^{k+1},
\end{align}
$a=\frac{\lambda_{te}\lambda_{sr}\xi}{ \lambda_{se}}$, $b=\frac{\lambda_{td}\lambda_{sr}\xi}{k\lambda_{sd}}$, 
$c_1=\frac{k\lambda_{sd}}{\Gamma_S}$, and $c_2=\frac{k\lambda_{sd}+\lambda_{se}}{\Gamma_S}$.
Finally, we use the partial fraction method from (\ref{PF_ERG_OTS_I_1}) and (\ref{PF_ERG_OTS_I_2}) to obtain the result in
(\ref{eq_asym_erg_ots}).
\end{proof}
\end{corollary}
\subsubsection{Minimal Interference Selection (MIS)}
Following Corollary \ref{ASY_ERG_COR}, the asymptotic ergodic secrecy capacity for this scheme can be derived as
\begin{align}
\mathcal{C}_{erg}
=&A ab\left(\frac{ln(a)}{(a-1)(a-b)}+\frac{ln(b)}{(b-1)(b-a)}\right)
\xd
-&A b\frac{ln(b)}{1-b},
\end{align}
where $A=\frac{\Lambda}{ln(2)}$, $a=\frac{\lambda_{te}\lambda_{sr}\xi}{ \lambda_{se}}$, and
$b=\frac{\lambda_{td}\lambda_{sr}\xi}{\lambda_{sd}}$.
\subsubsection{Minimal Eavesdropping Selection (MES)}
Similar to Corollary \ref{ASY_ERG_COR}, the asymptotic ergodic secrecy capacity for this scheme can be derived as
\begin{align}
\mathcal{C}_{erg}
=&A ab\left(\frac{ln(a)}{(a-1)(a-b)}+\frac{ln(b)}{(b-1)(b-a)}\right)
\xd
-&A b\frac{ln(b)}{1-b},
\end{align}
where $A=\frac{\Lambda}{ln(2)}$, $a=\frac{\lambda_{te}\lambda_{sr}\xi}{ K\lambda_{se}}$,  and 
$b=\frac{\lambda_{td}\lambda_{sr}\xi}{\lambda_{sd}}$. 
\section{Numerical Results and Discussions}\label{section 8}
In this section, theoretical results are plotted along with Monte Carlo simulations. 
Without loss of generality, we assume that all nodes are affected by the same noise power $N_0$, 
and the following parameters are set: $\beta$=0.5 bits/s/Hz, $R_{th}=0.5$ bits/s/Hz, 
$\lbrace\lambda_{tr}, \lambda_{td}, \lambda_{sd}, \lambda_{sr}, 
\lambda_{te}\rbrace$ = $\lbrace 3, -6, 3, -3, 6\rbrace$ dB. 
 {We use the Mathematica software to evaluate the integrals in (\ref{non_zero_eqt_OS}) and (\ref{secrecy OP equation OS}) for the non-zero secrecy rate and SOP of OS scheme, respectively.}
\subsection{Non-zero Secrecy Rate}
\begin{figure}
 \centering
 \includegraphics[width=2.8in]{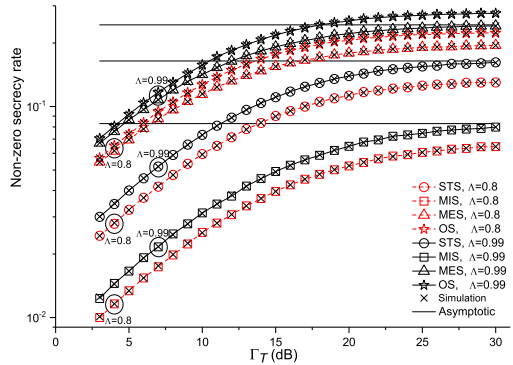} 
  \vspace*{-.5cm}
 \caption{
 Non-zero secrecy rate versus $\Gamma_T$(dB) for different values of $\Lambda$.}
 \label{fig:NonZerovaryS}
 \end{figure}

Fig. $\ref{fig:NonZerovaryS}$ plots the non-zero secrecy rate versus $\Gamma_T$ for 
two different values of the backhaul reliability, $\Lambda=0.8$ and $\Lambda=0.99$. The network 
parameters are set as $\lambda_{se}=30$ dB, $\textit{K}$=6 and $\Phi=0.1$. 
 {We observe that the analytical results perfectly match simulation results for both optimal and sub-optimal selections.
The performance from the best to the worst is: OS, MES, STS, and MIS.} We observe that the increase in $\Gamma_T$ improves the non-zero 
secrecy rate and reaches to its asymptotic limit. As $\Gamma_T$ increases, the transmit power of the small cell transmitter (i.e. secondary user) is permitted to increase as well, which leads to this improvement in performance. Eventually, the 
transmit power of the secondary user is restricted by the primary network outage 
probability, and hence, the performance reaches its asymptotic value. An increase in the value of $\Lambda$ can also increase the non-zero secrecy rate. It is intuitive to understand that increasing the reliability of the backhaul link improves the non-zero secrecy rate.
 
\begin{figure}
 \centering 
 \includegraphics[width=2.8in]{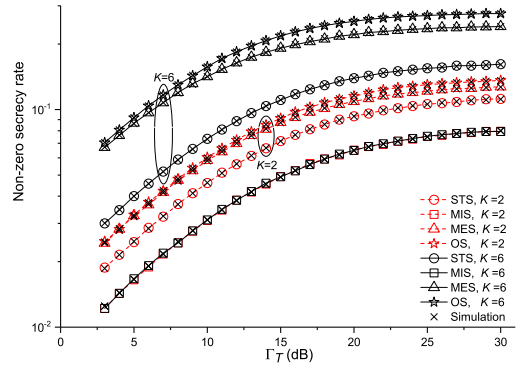} 
 \vspace*{-.5cm}
 \caption{
 Non-zero secrecy rate versus $\Gamma_T$(dB) for different number of secondary users, $K$}
 \label{fig:NonZerovaryK}
 \end{figure}

Fig. $\ref{fig:NonZerovaryK}$ shows the non-zero secrecy rate versus $\Gamma_T$ for
different numbers of small-cell transmitters, $\textit{K}$=2 and $\textit{K}$=6. 
The network parameters are set as $\Phi=0.1$, $\Lambda=0.99$, and $\lambda_{se}=30$ dB. 
 {As the number of small-cell transmitters  increases, the performance improves for the OS, STS, and MES schemes.} However, no visible improvement is seen for the MIS scheme.
With the increase in $\textit{K}$, the secondary users have a higher probability to choose a 
better transmitting channel to $D$ or worse channel to $E$, and hence,  {the performance for the OS, STS, and MES schemes 
improves.} However, due to the already imposed constraint for the primary QoS, the 
performance of the MIS scheme cannot improve with the increase in the number of possible interference links with $\textit{K}$.
\begin{figure}
 \centering
 \includegraphics[width=2.8in]{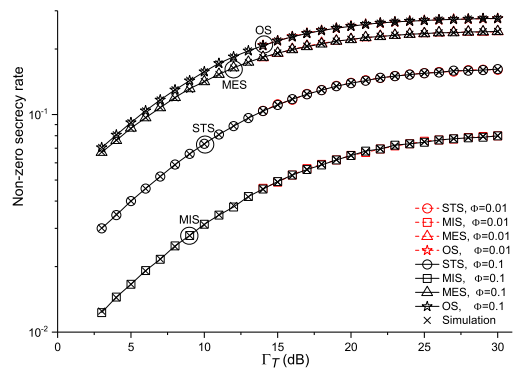} 
 \vspace*{-.5cm}
 \caption{
 Non-zero secrecy rate versus $\Gamma_T$(dB) for different values of $\Phi$.}
 \label{fig:NonZerovaryQoS}
 \end{figure}
 

In Fig. $\ref{fig:NonZerovaryQoS}$, the non-zero secrecy rate is investigated versus $\Gamma_T$ 
for two different values of the primary QoS constraint, $\Phi=0.01$, $\Phi=0.1$, with $\lambda_{se}=30$ dB, $K=6$, and $\Lambda=0.99$. As $\Gamma_T$ increases, performance saturates to its asymptotic value as in the the previous figures.
With the primary outage requirement $\Phi$ changes, no visible improvement in the performance can be seen. 
Non-zero secrecy rate depends on the probability that the legitimate channel is better than the illegitimate  channel, 
as an increase in $\Phi$ does not change this probability, the performance remains the same. In Fig. $\ref{fig:NonZerovaryQoS}$ we can see that the effect of increasing $\Phi$ is not as much as in Figs. $\ref{fig:SOPQoS}$ and $\ref{fig:EgodicQoS}$.


 \subsection{Secrecy Outage Probability}
 
\begin{figure}
 \centering
 \includegraphics[width=2.8in]{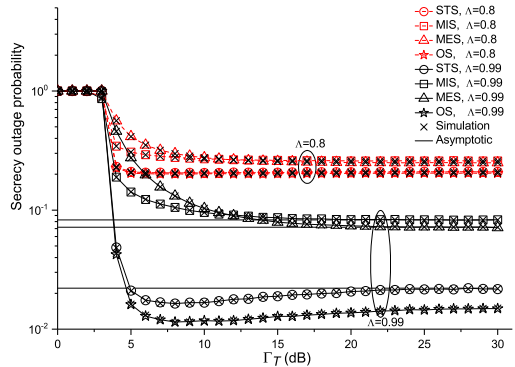} 
  \vspace*{-.5cm}
 \caption{
 SOP versus $\Gamma_T$(dB) for different values of $\Lambda$.}
 \label{fig:SOPvaryS}
 \end{figure}

Fig. $\ref{fig:SOPvaryS}$ plots the SOP versus $\Gamma_T$ for different values of the backhaul reliability, 
$\Lambda=0.8$ and $\Lambda=0.99$, with $\textit{K}$=6, $\Phi=0.1$. Analysis exactly matches simulation results. 
 {It is observed that the order of the performance from the best to the worst is: OS, STS, MES, and MIS.} An observation in SOP is that MIS and MES cross each other at a certain $\Gamma_T$, where MES is worse than MIS below that $\Gamma_T$. However, better than MIS beyond that $\Gamma_T$. 
Comparing the figures for non-zero secrecy rate and SOP, it is clear that MIS is the worst 
in both cases. In addition, MES performs better than STS for the non-zero secrecy rate and 
STS is better than MES in SOP. 
As in the non-zero secrecy rate case, SOP decreases when the reliability of the backhaul, $\Lambda$, increases. 
As $\Gamma_T$ increases, SOP reaches its floor, which is its asymptotic limiting value.
 
\begin{figure}
 \centering
 \includegraphics[width=2.8in]{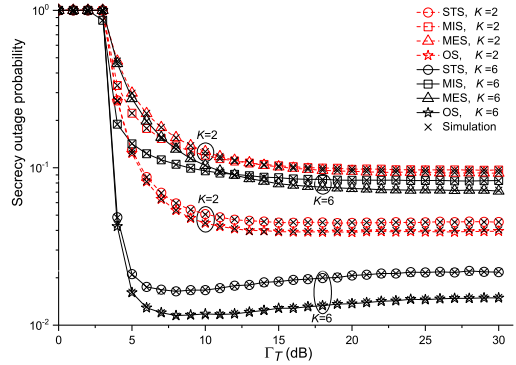} 
 \vspace*{-.5cm}
 \caption{
 SOP versus $\Gamma_T$(dB) for different number of secondary users, $K$}
 \label{fig:SOPvaryK}
 \end{figure}

Fig. $\ref{fig:SOPvaryK}$ shows the SOP versus $\Gamma_T$ for different numbers of small-cell 
transmitters, $\textit{K}$=2 and $\textit{K}$=6. The network parameters are set as 
$\Phi=0.1$ and $\Lambda=0.99$. As the number of small-cell transmitters $K$ increase, SOP improves for all schemes. This is in contrast to the non-zero secrecy rate case, as the MIS scheme benefits from the increase in $K$ as well. This concludes that as the number 
of transmitters increase, diversity increases for all schemes.

\begin{figure}
 \centering
 \includegraphics[width=2.8in]{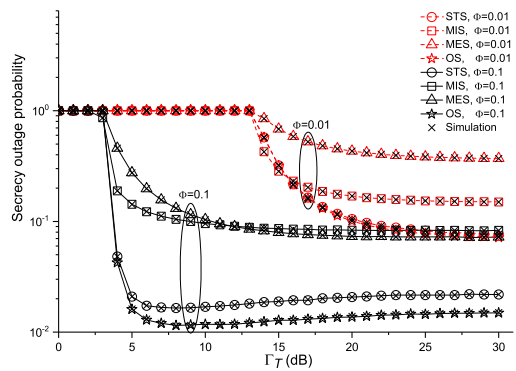} 
  \vspace*{-.5cm}
 \caption{
 SOP versus $\Gamma_T$(dB) for different values of $\Phi$.}
 \label{fig:SOPQoS}
 \end{figure}

Fig. $\ref{fig:SOPQoS}$ investigates SOP versus $\Gamma_T$ for two different values of the primary QoS constraint, 
$\Phi=0.01$ and $\Phi=0.1$ with $\textit{K}$=6, $\Lambda=0.99$. We observe that the increase in $\Phi$ results in a 
reduction in SOP. This is because the secondary network transmitters are allowed to have higher transmit power by 
relaxing the QoS requirement of the primary network. 
The figure also shows that the order of the performance from the best to the worst 
changes with $\Phi$.  {Although the OS and STS schemes always perform better than other two schemes with OS is the best}, the MES scheme is better than the MIS scheme at $\Phi=0.01$,  while the opposite is true at $\Phi=0.1$. Both of the above observations are in contrast 
to the non-zero secrecy rate case.
 
 
 
At relatively high $\Lambda$= 0.99, $K$=6, and $\Phi$=0.1, a particular observation from the SOP plots is that the SOP of the STS scheme initially goes below the asymptotic limit; however, it reaches its asymptotic value as $\Gamma_T$ increased further. This can only be seen for the curves plotted in black colour.

\subsection{Ergodic Secrecy Capacity}

\begin{figure}
 \centering
 \includegraphics[width=2.8in]{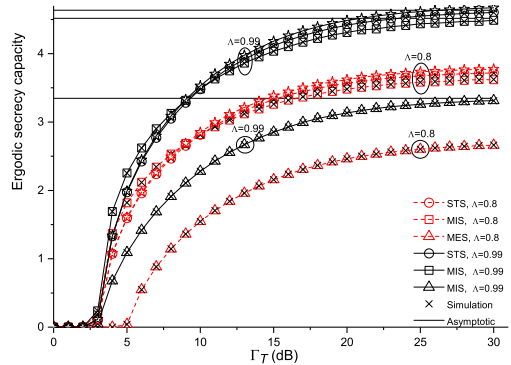} 
 \vspace*{-.5cm}
 \caption{
 Ergodic secrecy capacity versus $\Gamma_T$(dB) for different values of $\Lambda$.}
 \label{fig:EgodicvaryS}
 \end{figure}

Fig. $\ref{fig:EgodicvaryS}$ plots the ergodic secrecy capacity versus $\Gamma_T$ for two different values 
of $\Lambda$, $\Lambda=0.8$ and $\Lambda=0.99$. We set K=6, $\Phi=0.1$, and $\lambda_{se}=-3$ dB. 
As $\Gamma_T$ increases, the ergodic secrecy capacity increases and reaches its asymptotic value in high SINR regime.
It can be seen that the order of the performance of ergodic secrecy capacity from the best to 
the worst is: STS, MIS and MES. The order is different from both the non-zero secrecy rate and SOP.
The ergodic secrecy capacity increases when the reliability in the bakchaul link increases (i.e., when $\Lambda$ increases 
from 0.8 to 0.99, similar to the SOP case). 

\begin{figure}
 \centering
 \includegraphics[width=2.8in]{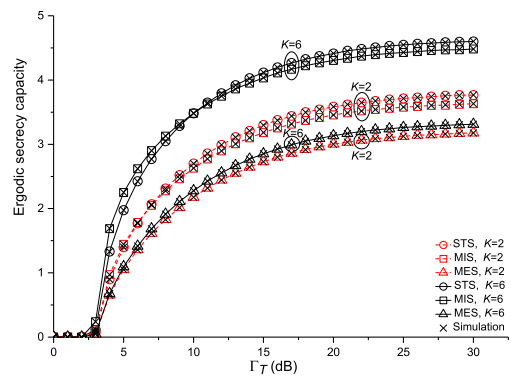} 
  \vspace*{-.5cm}
 \caption{
 Ergodic secrecy capacity versus $\Gamma_T$(dB) for different number of secondary users, $K$}
 \label{fig:EgodicvaryK}
 \end{figure}

Fig. $\ref{fig:EgodicvaryK}$ shows the ergodic secrecy capacity versus $\Gamma_T$ for different 
numbers of small-cell transmitters, $\textit{K}$=2 and $\textit{K}$=6. The network parameters 
are set as $\Phi=0.1$, $\Lambda=0.99$, and $\lambda_{se}=-3$ dB. 
It can be observed that the ergodic secrecy capacity increases as $\textit{K}$ increases.

\begin{figure}
 \centering
 \includegraphics[width=2.8in]{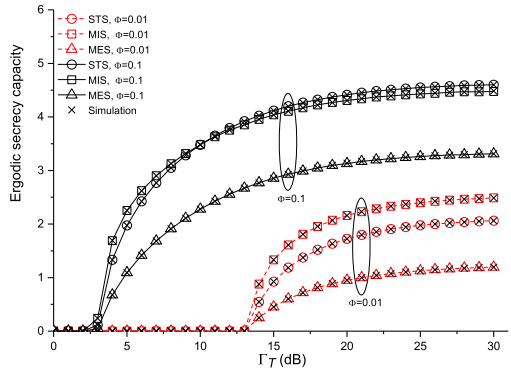} 
  \vspace*{-.5cm}
 \caption{
 Ergodic secrecy capacity versus $\Gamma_T$(dB) for different values of $\Phi$.}
 \label{fig:EgodicQoS}
 \end{figure}

In Fig. $\ref{fig:EgodicQoS}$, the ergodic secrecy capacity is investigated versus $\Gamma_T$ 
with different values of $\Phi$, $\Phi=0.01$ and $\Phi=0.1$. We set $\textit{K}$=6, 
$\Lambda=0.99$, and $\lambda_{se}=-3$ dB. We observe that an increase in $\Phi$ results in an increase in the ergodic secrecy capacity. Due to the increase in the QoS of the primary network, 
the secondary network is allowed to raise the transmit power, and hence, the previous observation. This observation is similar to the SOP case; however, the order of performance from the best to the worst selection scheme is different. 
 
 

A common observation from the SOP and ergodic secrecy rate plots is that the SOP remains 
constant to unity and ergodic secrecy rate remains constant to zero till $\Gamma_T$ reaches a certain value. 
This value actually depends on the primary QoS constraint, $\Phi$, which can be ascertained 
from the SOP and ergodic secrecy rate plots in   Figs. $\ref{fig:SOPQoS}$ and $\ref{fig:EgodicQoS}$, respectively.
This is because as $\Gamma_T$ is low, the transmit power allowed for the secondary transmitter at a given primary QoS constraint 
is also low, and cannot improve the secondary performance till $\Gamma_T$ allows it to reach a certain level. 

    \begin{figure}
 \centering
 \includegraphics[width=2.8in]{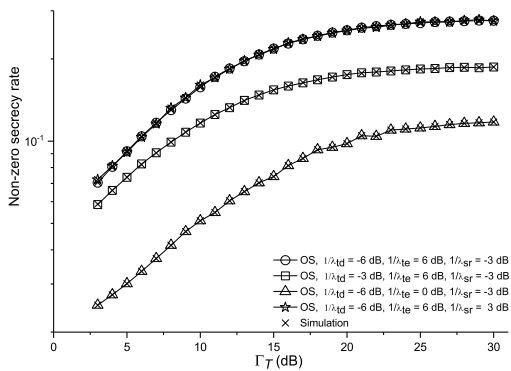} 
 \vspace*{-.5cm}
 \caption{
   {Non-zero secrecy rate versus $\Gamma_T$(dB) for different values of primary to secondary channel quality and vice versa.}}
 \label{fig:NonZerosnr}
 \end{figure}
 
  {To demonstrate the impact of the interference from the primary network to the secondary network and vice versa, we have shown secrecy performances by improving the channel quality of the primary to the secondary and the secondary to the primary network. Fig. \ref{fig:NonZerosnr} shows 
 the non-zero secrecy rate versus $\Gamma_T$ of the OS scheme for the three cases by i) improving  $1/\lambda_{td} = -6$ dB to $1/\lambda_{td} = -3$ dB for a given $1/\lambda_{te} = 6$ dB and $1/\lambda_{sr} = -3$ dB; ii) improving $1/\lambda_{te} = 0$ dB to  $1/\lambda_{te} = 6$ dB for a given $1/\lambda_{td} = -6$ dB and $1/\lambda_{sr} = -3$ dB; and iii) improving 
$1/\lambda_{sr} = -3$ dB to $1/\lambda_{sr} = 3$ dB for a given $1/\lambda_{td} = -6$ dB and $1/\lambda_{te} = 6$ dB, when $1/\lambda_{se}=30$ dB, $K = 6$, and $\Lambda = 0.99$. 
As the average channel quality of the link $T$-$D$ improves due to increase in $1/\lambda_{td}$, the secondary destination experiences more interference from the primary transmitter, and thus, the secrecy rate decreases. As $1/\lambda_{te}$ increases, the eavesdropper experiences more interference from the primary transmitter, and thus, the secrecy rate increases. As $1/\lambda_{sr}$ increases, no visible change in the performance can be seen. This can be explained, as the primary receiver experiences more interference from the secondary transmitter, and hence, the secondary transmit power automatically decreases due to primary QoS constraint; in turn, the non-zero secrecy rate remains the same.}

    \begin{figure}
 \centering
 \includegraphics[width=2.8in]{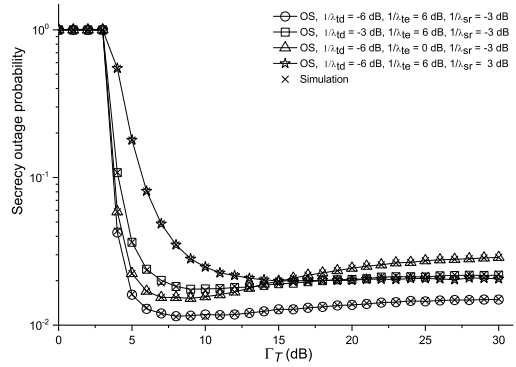} 
  \vspace*{-.5cm}
 \caption{
   {Secrecy outage probability versus $\Gamma_T$(dB) for different values of primary to secondary channel quality and vice versa.}}
 \label{fig:SOPsnr}
 \end{figure}

 {Fig. \ref{fig:SOPsnr} plots the SOP performance versus $\Gamma_T$ of the OS scheme for the same three cases as in the previous figure for non-zero secrecy rate, when $1/\lambda_{se} = -3$ dB, $K=6$, and $\Lambda = 0.99$. The conclusion is the same as in the previous figure with non-zero secrecy rate. The only difference is that when the $S$-$R$ channel quality improves, SOP increases in this figure, whereas, the non-zero secrecy rate did not change in the earlier figure. As the interference at $R$ increases, the secondary transmit power decreases, and as a result, the secrecy rate decreases and SOP increases. On the other hand, as the non-zero secrecy rate shows the probability of whether secrecy rate exceeds zero or not, it is possible to get a higher SOP when the non-zero secrecy rate remains the same.}
\section{Conclusion}\label{section 9}
 {Four optimal and sub-optimal transmitter selection schemes are proposed} to enhance the secrecy of a secondary small-cell cognitive radio network. 
The close-form expressions for the non-zero secrecy rate, secrecy outage probability, 
and ergodic secrecy capacity, as well as asymptotic expressions are  {derived for sub-optimal selection schemes. Computable expressions for the non-zero secrecy rate and secrecy outage probability of the optimal selection scheme are also derived.}  
The results show that the increase in the primary transmitter's power and in the number of small-cell transmitters 
can enhance the system's secrecy performance for all selection schemes. As the primary transmitter's power increases beyond a certain value, the secrecy performance converges to its 
asymptotic value. In addition, results show that the backhaul reliability and the desired 
outage probability at the primary user are important parameters for the secrecy 
performance. Increasing backhaul reliability  and relaxing the primary user QoS constraint improves the overall 
secrecy of the secondary system. 

\bibliographystyle{IEEEtran}
\ifCLASSOPTIONcaptionsoff
  \newpage
\fi

\begin{biography}[{\includegraphics[width=1in,height=1.25in,clip,keepaspectratio]{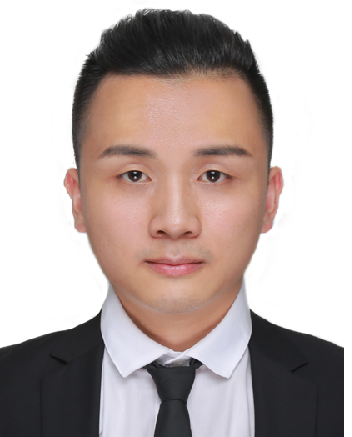}}]
{\bf Jinghua Zhang}  is currently a wireless product manager in Huawei at Reading, UK. He received his B.Eng. degree in Electronic and Electrical Engineering and Ph.D. degree in Wireless Communications from Queen's University Belfast in 2015 and 2019, respectively. His research interests include physical layer security, cognitive radio communication, and energy harvesting communications.
\end{biography}
\begin{biography}[{\includegraphics[width=1in,height=1.25in,clip,keepaspectratio]{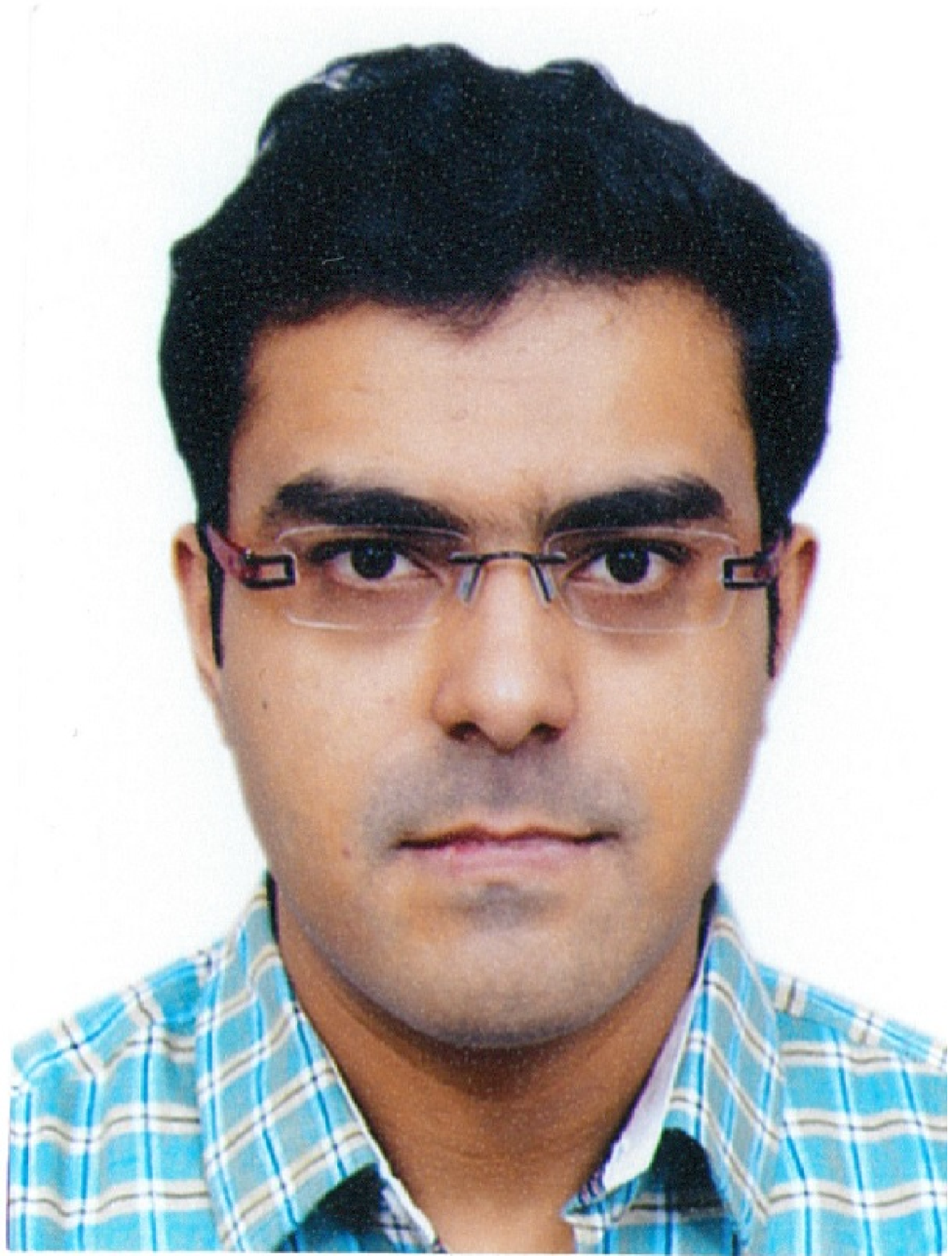}}]
{\bf Chinmoy Kundu} (S’12--M’15) received 
his Ph.D. degree in communication engineering from IIT Delhi in March, 2015. He is currently a DST Inspire Faculty at IIT Jammu. He was a Research Associate in the University of Texas at Dallas in 2018, Newton International Fellow in Queen's University Belfast from 2016 to 2018, and a Post-Doctoral fellow in Memorial University Canada during 2015 to 2016. He is a recipient of the Newton International Fellowship from the Royal Society, UK, Inspire Faculty award twice from the DST, Government of India, and Junior Research Fellowship from the CSIR, Government of India. His research interests include 5G enabling technologies, physical layer security, cognitive radio systems, optimization, and cooperative communications.
\end{biography}
\begin{biography}[{\includegraphics[width=1in,height=1.25in,clip,keepaspectratio]{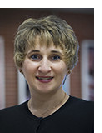}}]
{\bf Octavia A. Dobre} (M'05--SM'07) 
received the Dipl. Ing. and Ph.D. degrees from Politehnica University of Bucharest (formerly Polytechnic Institute of Bucharest), Romania, in 1991 and 2000, respectively. Between 2002 and 2005, she was with New Jersey Institute of Technology, USA and Politehnica University of Bucharest. In 2005, she joined Memorial University, Canada, where she is currently Professor and Research Chair. She was a Visiting Professor with Massachusetts Institute of Technology, USA and Université de Bretagne Occidentale, France.

Her research interests include enabling technologies for 5G and beyond, blind signal identification and parameter estimation techniques, as well as optical and underwater communications. She authored and co-authored over 250 refereed papers in these areas.

Dr. Dobre serves as the Editor-in-Chief (EiC) of the IEEE Open Journal of the Communications Society, as well as an Editor of the IEEE Communications Surveys and Tutorials, IEEE Vehicular Communications Magazine, and IEEE Systems. She was the EiC of the IEEE Communications Letters, as well as Senior Editor, Editor, and Guest Editor for various prestigious journals and magazines.  Dr. Dobre was the General Chair, Technical Program Co-Chair, Tutorial Co-Chair, and Technical Co-Chair of symposia at numerous conferences, including IEEE ICC and IEEE Globecom. 

Dr. Dobre was a Royal Society Scholar in 2000 and a Fulbright Scholar in 2001. She obtained Best Paper Awards as various conferences, including IEEE ICC and IEEE WCNC. Dr. Dobre is a Distinguished Lecturer of the IEEE Communications Society and a Fellow of the Engineering Institute of Canada. 
\end{biography}
\begin{biography}[{\includegraphics[width=1in,height=1.25in,clip,keepaspectratio]{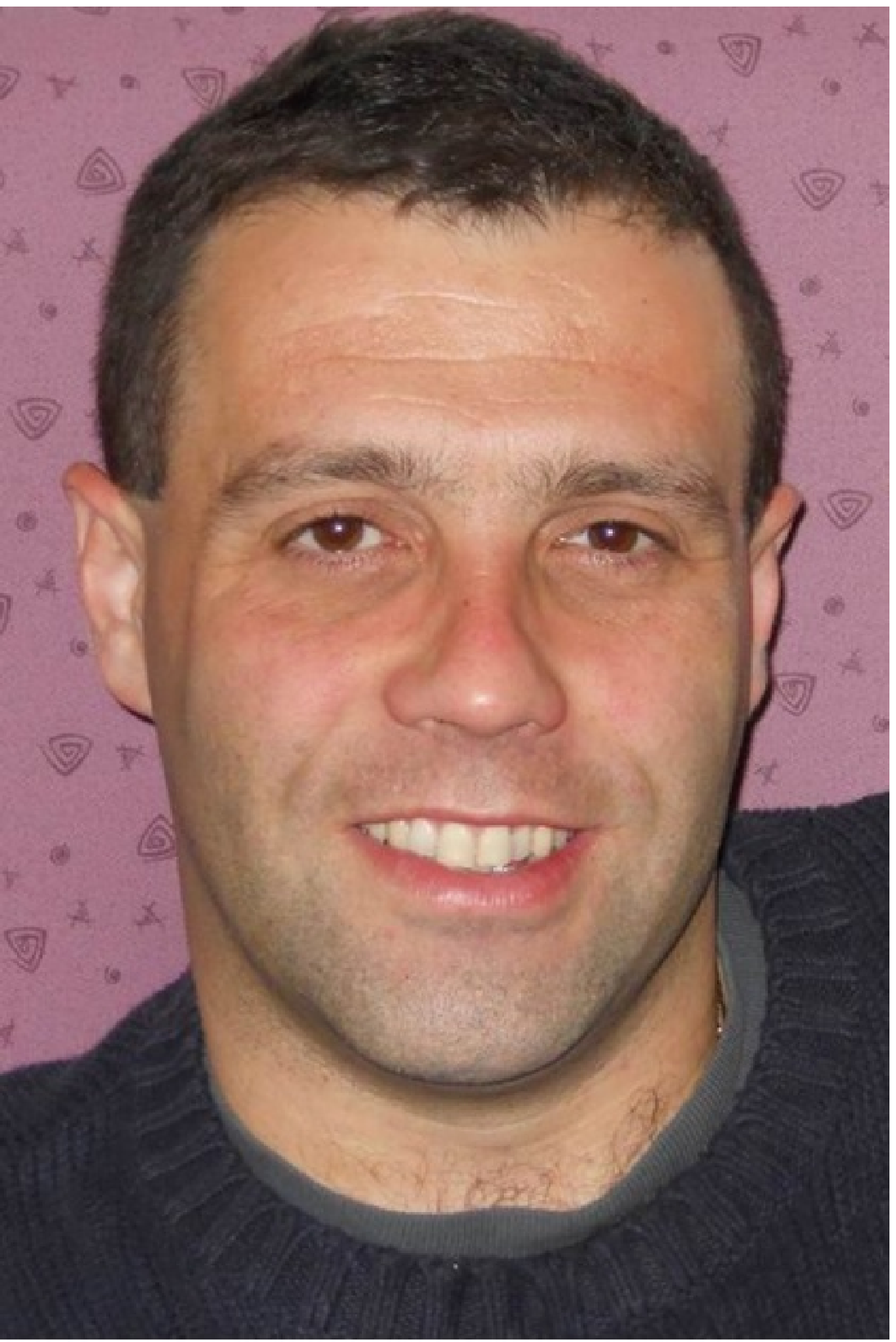}}]
{\bf Emiliano Garcia-Palacios } is a Lecturer at the School of Electrical, Electronic Engineering and Computer Science, Queen's University Belfast. He received his Ph.D from Queen’s University in 2000 and since then he has been leading research in wireless networks resource management. His research interests include wireless resource allocation and optimization for 5G, secure networks, sensor networks and IoT.
\end{biography}
\begin{biography}[{\includegraphics[width=1in,height=1.25in,clip,keepaspectratio]{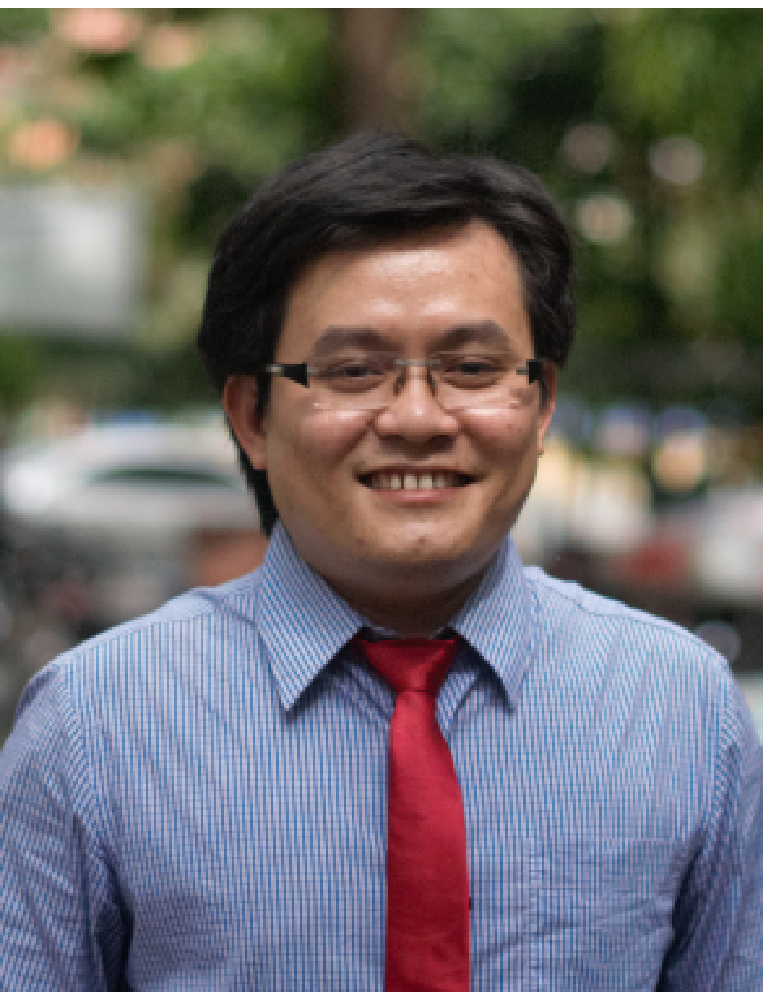}}]
{\bf Nguyen-Son Vo } Nguyen-Son Vo received the Ph.D. degree in communication and information systems from Huazhong University of Science and Technology, China, in 2012. He is with the Institute of Fundamental and Applied Sciences, Duy Tan University, Ho Chi Minh City, Vietnam. His research interests focus on self-powered multimedia wireless communications, quality of experience provision in wireless networks for smart cities, IoT to disaster and environment management.
He received the Best Paper Award at the IEEE Global Communications Conference 2016 and the prestigious Newton Prize 2017. 
He has been serving as an Associate Editor of IEEE Communications Letters since 2019.
\end{biography}
\end{document}